\documentclass[aps,pra,showpacs,preprintnumbers,preprint,a4paper]{revtex4}
\usepackage{amssymb,amsmath}
\usepackage{dcolumn}
\usepackage{bm}
\usepackage{graphicx}

\begin{document}

\title{A quantitative theory-versus-experiment comparison for the intense laser dissociation of H$_2^+$.}

\author{V.N.~Serov}
\altaffiliation[Also at ]{Institute of Physics, St.Petersburg State University \\
                          Peterhof, St.Petersburg, 198504 Russia}
\email{vassili.serov@ppm.u-psud.fr}
\author{A.~Keller}
\affiliation{Laboratoire de Photophysique Mol\'eculaire du CNRS,Universit\'e de Paris-Sud, 91405 Orsay, France}
\author{N.~Billy}
\affiliation{Laboratoire Kastler Brossel, Universit\'e d'Evry Val d'Essonne, Boulevard Fran\c{c}ois Mitterrand, 91025 Evry cedex, France}
\author{O.~Atabek}
\affiliation{Laboratoire de Photophysique Mol\'eculaire du CNRS,Universit\'e de Paris-Sud, 91405 Orsay, France}

\date{\today}

\begin{abstract}

A detailed theory-versus-experiment comparison is worked out for H$_2^+$ intense laser 
dissociation, based on angularly resolved photodissociation spectra recently recorded in H.Figger's 
group. As opposite to other experimental setups, it is an electric discharge (and not an optical 
excitation) that prepares the molecular ion, with the advantage for the theoretical approach, to neglect without lost of accuracy,
the otherwise important ionization-dissociation competition. Abel 
transformation relates the dissociation probability starting from a single ro-vibrational state, to the 
probability of observing a hydrogen atom at a given pixel of the detector plate. Some statistics on initial 
ro-vibrational distributions, together with a spatial averaging over laser focus area, lead to 
photofragments kinetic spectra, with well separated peaks attributed to single vibrational levels.
An excellent theory-versus-experiment agreement is reached not only for the kinetic spectra, but also 
for the angular distributions of fragments originating from two different vibrational levels
resulting into more or less alignment. Some characteristic features can be interpreted in terms of basic 
mechanisms such as bond softening or vibrational trapping.

\end{abstract}


\maketitle

\section{Introduction}

The above threshold multiphoton ionization and dissociation of H$_2^+$ subjected to strong laser
interaction, have revealed interesting nonlinear effects in angularly resolved kinetic energy distributions
of the photofragments, measured in experimental works covering the last decade \cite{1,2,3,4,5}. Among these
are the observations of  very large increase (or sometimes decrease) of the photodissociation rates
originating from some vibrational states of the parent molecule at some specific laser intensities or,
even more unexpectedly, misalignment effects in fragments angular distributions \cite{6}.
The interpretation of such behaviors has been attempted by referring to some basic dynamical
mechanisms evidenced through the light-induced adiabatic potentials describing the dressed states of the molecule-plus-field system.
According to the frequency regimes, bond softening (in UV) \cite{1,7}
or barrier suppression (in IR) \cite{8} mechanisms tend to enhance the dissociation cross-section especially in the polarization
direction of the laser. As opposite to them vibrational trapping (in UV) \cite{9} or dynamical dissociation quenching (in IR) \cite{10},
act as stabilization mechanisms, favoring misalignment in the fragments distributions. This complementarity has also been referred to,
for laser control purposes of the chemical reactivity; namely by softening some bonds while hardening others \cite{11}.
Although very accurate quantum calculations in the frame of time dependent approaches have been carried out, with successful interpretations
of dynamical behaviors in short-intense laser pulses,
to the best of our knowledge, there is no a thorough and quantitative theory-versus-experiment comparison, up to date,
the work of Kondorskiy et al. \cite{Kondorskiy} being a precursor in this direction.
Basically two reasons can  be invoked for the difficulty of such an attempt: only very few theoretical models take into account
the competition between ionization and dissociation processes leading, in very strong fields, to Coulomb explosions
and only very few experimental works are conducted with a careful investigation of vibrational populations and sufficiently high
momentum and angular resolution yielding accurate information about the dissociation
of single vibrational levels.

Experimental works on this system can be classified according to the preparation of the parent ion
H$_2^+$ from the neutral molecule H$_2$. A first category collects experiments referring to optical ionization with a laser 
prepulse \cite{1,2,3,4}.
The independence of the ionization and dissociation processes can not be experimentally controlled, and their competition is still an
open question \cite{5}. More recently, another
kind of approach has been investigated through ion beam experiments, where H$_2^+$ ions are produced in a 
dc electric or plasma discharge that disentangle ionization and dissociation processes \cite{12,13}.
An accelerated and strongly collimated monochromatic H$_2^+$ beam is crossed at right angle by a focused intense laser beam. 
An advantage of the strong ion beam collimation is the reduction of the intensity volume effect; all ions being approximately
irradiated by the same laser intensity (the validity of such approximation will however be discussed hereafter).
Moreover, experiments conducted with low intensity pulses coupled to computational simulations of the resulting
dissociation spectra, allow the determination of the population of the rovibrational levels of H$_2^+$ molecules in the beam. The neutral
dissociation fragments (H atoms originating from photodissociation of H$_2^+$) are projected on a multichannel
detector (MCD), whereas the charged particles (undissociated H$_2^+$ molecules and H$^+$
fragments)  are extracted by deflection into a Faraday cup using an electric field. Excellent energy resolution (about 1\%) allows
the separation, in the circularly shaped patterns observed on the screen, the momentum projection of fragments almost originating
from a single vibrational level \cite{12}.

A model aiming in a quantitative theory-versus-experiment comparison, within the frame of the ion beam setup, has to fulfill the following requirements:

i) the photodissociation process has to be accurately described in the center of mass frame by a wavepacket
propagation under the effect of an intense radiative field, starting from a given rovibrational state. There is no need,
however, to refer to any competition with ionization, as the experiment precisely disentangles these two fragmentation processes.

ii) a geometrical transformation towards the MCD-plate has to
be carried out, taking into account the macroscopic kinetics of the ion beam. This relates the total number of particles collected by a
given pixel of the plate, during the whole experiment, to the
previously calculated wavepacket, describing the evolution of an initial rovibrational state under the
effect of a laser pulse of a given intensity.

iii) although particular attention has been paid to the ion beam collimation in order to reduce the field intensity volume effects, a spatial
average over the laser focusing area has to be carried, taking into account the different radiative couplings felt by  H$_2^+$ molecules
according to their geometrical position in the beam. This can be done through the use of some experimental measurements of the intensity
distribution in the focus carried through a pinhole of 1 $\mu$m diameter \cite{12}.

iv) quantitative agreement also requires an averaging on the detector plate using some windowing functions that simulate 
the resolution power of the detector.

The organization of the paper follows these achievements in Section~\ref{sec:Theory}. The results and their
interpretation are presented in Section~\ref{sec:Results} with a thorough discussion of the role of the intensity volume effect.
An excellent theory-versus-experiment agreement is obtained not only for the kinetic but also on the angular distributions
of the photofragments. Section IV is devoted to some conclusions and perspectives.

\section{Theory}
\label{sec:Theory}

Referring only to two radiatively coupled Born-Oppenheimer electronic states; namely the ground
(1s$\sigma_g$) and the first excited (2p$\sigma_u$), an accurate wavepacket propagation method using 
the split operator technique is described in detail in ref.\cite{14,15}.
For the sake of completeness, we give hereafter a brief summary of the method, introducing the corresponding coordinates,
operators and quantum numbers. The emphasis is rather put on the way to relate the quantum information content of the wavepacket to the
observed momentum projections of the neutral photofragments H resulting from a rovibrational distribution of parent ions H$_2^+$ excited
by a laser source of given spatial distribution.

\subsection{The wavepacket propagation}
In the laboratory frame and using spherical coordinates, the total molecule-plus-field Hamiltonian
is written in terms of a two-by-two operator matrix:
\begin{equation}\label{eq:H_tot}
\mathbf{H}(R,\theta,\phi;t) =\mathbf{T}_R+\mathbf{T}_\theta+\mathbf{T}_\phi+\mathbf{V}(t).
\end{equation}
$\pmb{R}$ is the diatomic internuclear vector. $R$, $\theta$ and $\phi$ designate the internuclear distance, polar and azimuthal angles of $\pmb{R}$
with respect to the laser polarization vector $\pmb{\epsilon}$, respectively.
As is usually done, a functional change on the wavepacket:
\begin{equation}\label{eq:DefPhi}
\pmb{\Psi}(R,\theta,\phi;t)=\frac{1}{R} \pmb {\Phi} (R,\theta,\phi;t)
\end{equation}
aiming in a simplification of the radial part of the kinetic operators, leads to:
\begin{subequations}\label{eq:T_all}
\begin{equation}\label{eq:T_R}
\mathbf{T}_R=-\mathbf{1}\frac{1}{2 {\cal M}}\frac{\partial^2}{\partial R^2};
\end{equation}
\begin{equation}\label{eq:T_th}
\mathbf{T}_\theta=-\mathbf{1}\frac{1}{2 {\cal M} R^2}\frac{1}{\sin \theta} \frac{\partial}{\partial \theta}\left(\sin \theta \frac{\partial}{\partial \theta}\right);
\end{equation}
\begin{equation}\label{eq:T_phi}
\mathbf{T}_\phi=-\mathbf{1}\frac{1}{2 {\cal M} R^2}\frac{1}{\sin^2 \theta}\frac{\partial^2}{\partial\phi^2}
\end{equation}
\end{subequations}
with $\mathbf{1}$ the identity (2$\times$2) operator matrix. Atomic units ($\hbar$=1) are used in Eqs(\ref{eq:T_all})
where ${\cal M}$ designates the reduced mass. 
The time dependence arises in the non-diagonal terms of the potential energy operator matrix $\mathbf{V}$
through the radiative couplings:
\begin{equation}\label{eq:rad_coupl}
V_{12}(R,\theta,t)= \mu(R) \mathcal{E}(t) \cos \theta,
\end{equation}
where $\mu(R)$ is the transition dipole moment and $\mathcal{E}(t)$ is the laser electric field amplitude,
given as product of a pulse shape $\epsilon(t)$ times an oscillatory term involving the carrier wave frequency $\omega$:
\begin{equation}\label{eq:field}
\mathcal{E}(t)= \epsilon(t) \cos \omega t.
\end{equation}
Note that the $\cos \theta$ in Eq.(\ref{eq:rad_coupl}) results from the dot product of the transition dipole vector
(parallel to $\pmb{R}$) times the laser polarization vector $\pmb{\epsilon}$.

The diagonal elements $V_1(R)$ and $V_2(R)$ of $\pmb{V}$ are nothing but the BO curves of the ground (label 1)
and first excited (label 2) states of H$_2^+$. $V_1$, $V_2$ and $\mu$ are 
obtained in the frame of the Born-Oppenheimer approximation, at the zero order
level with respect to the ratio $m_e/m$ of the electron to the proton masses.
Using spheroidal coordinates, it is well known that the Schr\"odinger equation can
be written as two eigenvalue equations \cite{16a,16b}, which have been numerically solved
here using the shooting method \cite{16c}. The potential energy curves have been computed
in the range 0 $<R<$ 200 a.u., with a numerical accuracy checked to be better
than 10$^{-12}$ a.u. The mass ratio $m/m_e$ has been taken as $m/m_e$ = 1836.152701.
Finally the dipole matrix element $\mu$ between the $1s\sigma_g$ and $2p\sigma_u$
states has been obtained by numerical integration of the wave functions, at the
same level of numerical accuracy.

The time dependent Schr\"odinger equation (TDSE) describing the wavepacket propagation is:
\begin{equation} \label{eq:TDSE_1}
i\frac{\partial}{\partial t}\pmb{\Phi}(R,\theta,\phi;t)=\mathbf{H}(R,\theta,\phi;t)\pmb{\Phi}(R,\theta,\phi;t)
\end{equation}
with, as an initial condition:
\begin{equation}\label{eq:phi_vec0}
\pmb{\Phi}(R,\theta,\phi;t=0)=
\left( \begin{array}{c}
\Phi_1(R,\theta,\phi;0)\\
0
\end{array} \right)
\end{equation}
reflecting the fact that at time $t=0$, only the ro-vibrational levels of the ground electronic state are populated.
The eigenfunction $\Phi_1$ precisely corresponds to such a state with quantum numbers
$g$, $v$, $N$, $M_N$ (electronic ground, vibrational, total and $\pmb{\epsilon}$-projected rotational) and is given by:
\begin{equation}\label{eq:Phi0_chi}
\Phi_1(R,\theta,\phi;0)=\chi_{g,v,N}(R)P_N^{M_N}(\cos \theta)e^{i M_N \phi}.
\end{equation}
$P_N^{M_N}(\cos \theta)$ is the ($N$,$M_N$) Legendre polynomial, whereas the radial part is defined as the solution
of the time-independent Schr\"odinger equation:
\begin{equation}\label{eq:radSE}
\left[-\frac{1}{2 {\cal M}} \frac{{\mathrm d}^2}{{\mathrm d} R^2}+V_1(R)+\frac{N(N+1)}{2 {\cal M} R^2}-E_{v,N}\right]\chi_{g,v,N}(E)=0.
\end{equation}
The motion associated with the azimuthal angle $\phi$ remains separated under the action of the $\phi$-independent
$\pmb{V}$, such that $M_N$ is a good quantum number describing  the invariance through rotation about $\pmb{\epsilon}$.

The propagation using the split-operator technique has been described in full detail in previous works \cite{14,15,17}.
The  peculiarity of odd-charged homonuclear ions is their linearly increasing dipole moment with R, leading to
asymptotically divergent radiative couplings. We take them into account by splitting the wavefunction into two regions,
an internal and an asymptotic one. The latter is analyzed by a generalization of the Volkov type solutions
\cite{18}, while the numerical propagation on the former is performed by Fourier transform methodology \cite{19}
with the implementation of a unitary Cayley scheme for $T_\theta$ \cite{17}.

\subsection{From wavepacket to observed spectra}

The main concern of this paragraph is to relate the experimental observable, {\it i.e.} the probability distribution 
of hydrogen atoms resulting from H$_2^+$
photodissociation, as recorded on the multichannel detector (MCD), to the asymptotic part of the wavepacket
$\pmb{\Phi}(R,\theta,\phi;t)$ solution of Eq.(\ref{eq:TDSE_1}).
By asymptotic we mean large internuclear distances $R$ for which the molecule is considered as dissociated
without the possibility of a recombination process. To the best of our knowledge such a correlation has not rigorously been attempted
in the literature. So far, the interpretation of general tendencies of photodissociation spectra referring to basic mechanisms,
has rather been conducted by angularly resolved kinetic energy distribution given by:
\begin{equation}\label{eq:Pkth}
{\cal P}(k,\theta,\phi)=\lim_{t\to\infty}\left| \hat \Phi(k,\theta,\phi;t) \right|^2,
\end{equation}
where
\begin{equation} \label{eq:FFT}
\hat\Phi(k,\theta,\phi;t)=\frac{1}{\sqrt{2\pi}}\int_{-\infty}^{\infty} \Phi(R,\theta,\phi;t)e^{-ikR} {\mathrm d}R.
\end{equation}
is the Fourier transform of $\Phi$ over the scalar variable $R$, ({\it i.e.} not over $\pmb{R}$ taken as a vector).
The argument retained by doing so, is that asymptotically, due to $R^{-1}$ type of behavior in the kinetic operators Eqs(\ref{eq:T_th},\ref{eq:T_phi}),
angular dynamics is not affected at large internuclear distances. Note that in this paragraph, for the sake of simplicity,
we drop the labels of $\pmb{\Phi}$ depicting initial state quantum numbers ($v$, $N$, $M_N$).

\begin{figure}[!ht]
\begin{center}
   \includegraphics[scale=0.4,angle=0]{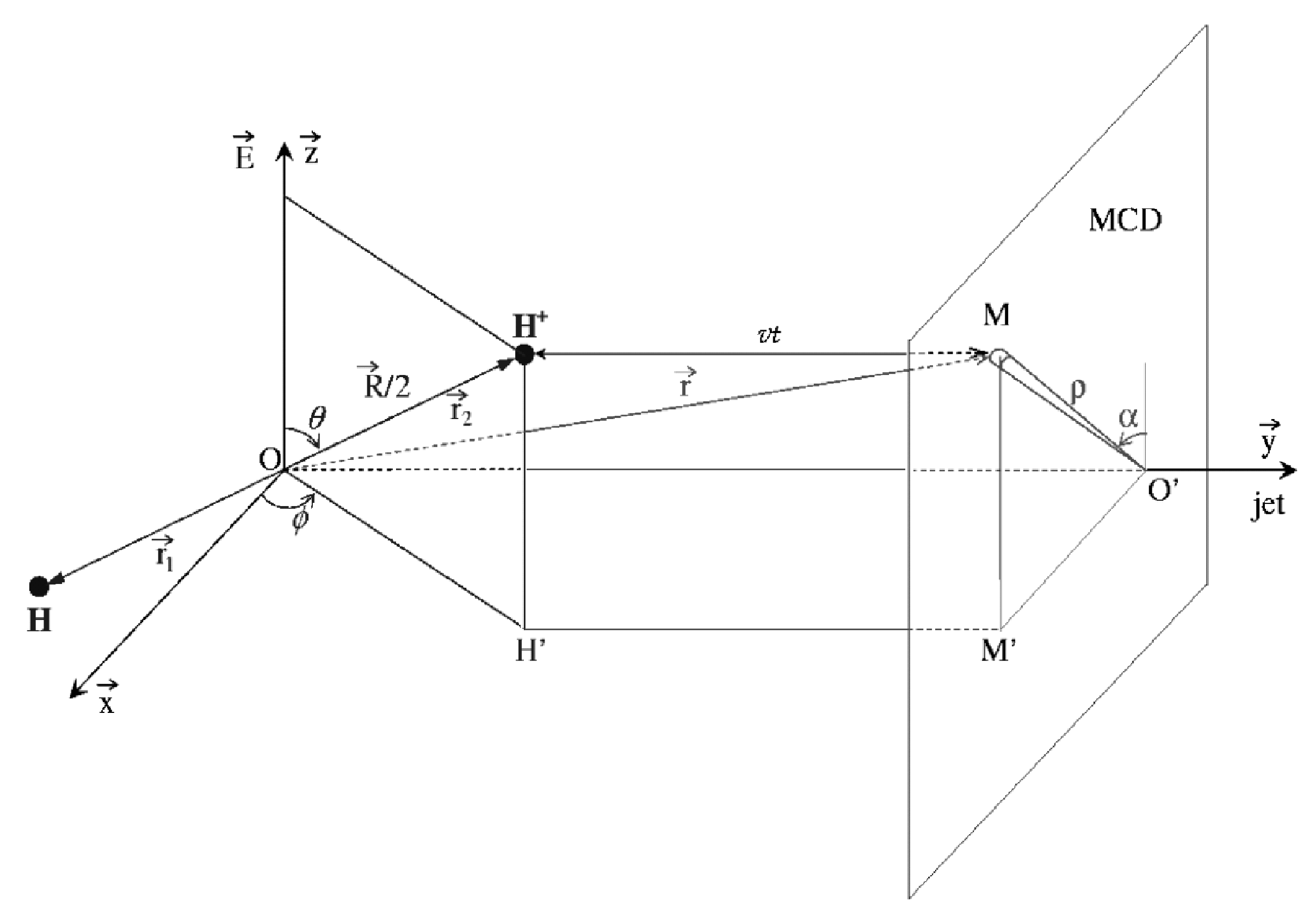}
\end{center}
\caption{\footnotesize The H$_2^+$ photodissociation experiment through the H$_2$ photoionization}\label{fig:shexp_1}
\end{figure}

To reach a comparative level of understanding, we are now describing the two families of experiments. Pertaining to the first family are
photodissociation experiments where both photodissociation and photoionization steps are laser induced \cite{1,2,3,4}.
Starting from neutral H$_2$, in its ground electronic and vibrationless state X($v$=0), a multiphoton excitation leads, through the EF
intermediate excited electronic state, to the H$_2^+$ ground state with a distribution of rovibrational levels. Dissociation follows
the absorption of additional photons and is very fast as compared to the relative motion of the parent ion H$_2^+$ in the laboratory frame.
Whence the photofragments are well separated, H$^+$ ions are extracted (accelerated) through an electric field and collected on the MCD plate.
A schematic view is provided in figure \ref{fig:shexp_1}. Photodissociation occurs, as a fast process, at the origin $O$ of the laboratory frame,
at a time which is taken as $t=0$. The laser polarization vector is along the $z$-direction, $\pmb{r_1}$ and $\pmb{r_2}$ are the vectors
pointing H and H$^+$. A further step is the extraction of the proton H$^+$ by an electric field applied along the $\pmb{y}$-direction towards
the MCD plate positioned at a distance $OO'$=$D$ from the origin. The detection occurs on a pixel $M$ defined by its polar coordinates ($\rho$, $\alpha$)
on the MCD surface (or by $\pmb{r}$ with respect to $O$) that H$^+$ is reaching after a time of flight $t$, with velocity $v$. It is worthwhile noting
that this last step is just a mapping of the photofragment onto the detector (without dissociation during time $t$).
The vector transformation relating the proton H$^+$ position ($R,\theta,\phi$) in the center of mass frame to the pixel 
M~($\rho, \alpha$) on the detector is known as the Abel transformation \cite{23}.

\begin{figure}[!ht]
\begin{center}
   \includegraphics[scale=0.4,angle=0]{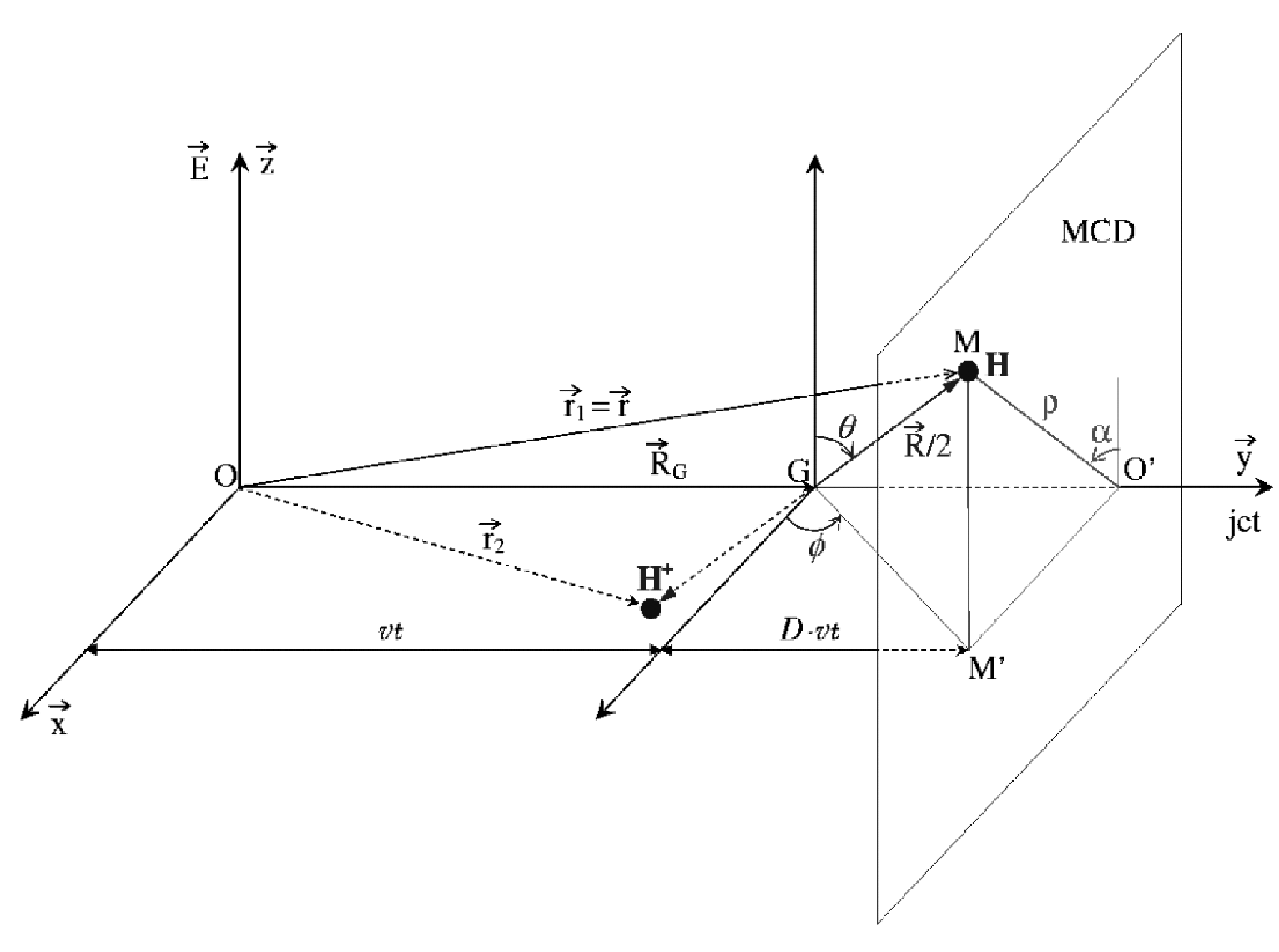}
\end{center}
\caption{\footnotesize The H$_2^+$ photodissociation experiment based on the ionization of H$_2$ using discharge source}\label{fig:shexp_2}
\end{figure}

A different situation prevails in the experiments of the second family where an electric 
or a plasma discharge ionizes H$_2$ into H$_2^+$ \cite{12,13}.
The resulting ion beam is strongly accelerated by an electric field and is crossed at $t=0$ by the laser beam
at a point $O$ of the laboratory frame. 
The description of such experiments, as illustrated in figure \ref{fig:shexp_2}, has to combine two motions; namely, the
translation of the center of mass $G$ in the laboratory frame along $\pmb{u_y}$ (unit vector along $\pmb{y}$) with velocity $v$ and the nuclear separation
(dissociation) in the center of mass frame.
The hydrogen atom H resulting from photofragmentation is collected at the pixel $M$ of the detector. It is to be noted that
$M$ is positioned with respect to the laboratory frame with a vector $\pmb{r}$, corresponding to $\pmb{r_1}$
at time $t$, when H reaches $M$.

As, our concern is the quantitative interpretation of photodissociation spectra obtained in H.Figger's group
using an electric discharge to induce ionization \cite{12,20,21}, emphasis is put in the following on a thorough description
of the kinematics of the second family experiments. The quantity which is measured, is nothing but the number of hydrogen atoms ${\mathrm d} N$
collected on each pixel $M$~($\rho$, $\alpha$) at infinite time. This can ultimately been related with the time integral of the flux of the
current density $\pmb{j}(\rho,\alpha,t)$ of H orthogonal to the area ${\mathrm d} S= \rho {\mathrm d}\rho {\mathrm d}\alpha$ of the finite size pixel $M$, as
\begin{equation}\label{eq:DN}
{\mathrm d}N(\rho,\alpha)=N {\mathrm d}S\int_0^{\infty} \pmb{j}(\rho, \alpha,t) \cdot \pmb{u_y}{\mathrm d}t,
\end{equation}
where $N$ is the total number of H photofragments.
The flux in Eq.(\ref{eq:DN}), involves an averaging over the positions of all protons H$^+$ that are not detected in the experiment
\cite{125}:
\begin{equation}\label{eq:j1}
\pmb{j}(t)=\frac{1}{m}\int {\mathrm d}\pmb{r_2}\  \mathrm{Im} \left[\Psi^*(\pmb{R},\pmb{R_G};t) \nabla_{\pmb{r_1}} \Psi(\pmb{R},\pmb{R_G};t)\right].
\end{equation}
where $\Psi(\pmb{R},\pmb{R_G};t)$ is the overwhole wavepacket describing the combined molecular internal $\pmb{R}$-motion
and center of mass $\pmb{R_G}$-motion. $\mathrm{Im}$ stands for the imaginary part. Frame transformations 
defining $\pmb{R_G}$ and some vector relations directly related with figure \ref{fig:shexp_2},
are gathered in an Appendix.
Separation of the photofragments relative motion described by $\Psi(\pmb{R};t)$, from the motion of the center of mass
described by $\Phi_G(\pmb{R_G};t)$, leads to the following representation of the total wavefunction:
\begin{equation}\label{eq:Psi_tot}
\Psi(\pmb{R},\pmb{R_G};t)=\Psi(\pmb{R};t)\Phi_G(\pmb{R_G};t).
\end{equation}
Using the frame transformations Eqs(\ref{eq:A1},\ref{eq:A2}) together with Eq.(\ref{eq:DefPhi}) one has:
\begin{equation}\label{eq:Psi_tot_2}
\Psi(\pmb{R},\pmb{R_G};t)=\frac{\Phi(\pmb{r_1}-\pmb{r_2};t)}{||\pmb{r_1}-\pmb{r_2}||}\Phi_G\left(\frac{(\pmb{r_1}+\pmb{r_2})}{2};t\right).
\end{equation}
Concerning the calculation of the gradient $\nabla_{\pmb{r_1}}$ involved in Eq.(\ref{eq:j1}),
we note that the flux has to be evaluated at large $R$ with as consequences:
\begin{equation}\label{eq:nabla_r1Phi}
\nabla_{\pmb{r_1}}\Phi(\pmb{R};t) = \nabla_{\pmb{R}}\Phi(\pmb{R};t)\simeq \pmb{u_R} \frac{\partial}{\partial R} \Phi(\pmb{R};t)
\end{equation}
with $\pmb{u_R}$ the unit vector along $\pmb{R}$ (Eq.(\ref{eq:A6}))
and
\begin{equation}\label{eq:nabla_r1PhiG}
\nabla_{\pmb{r_1}}\Phi_G(\pmb{R_G};t)=\frac{1}{2}\nabla_{\pmb{R_G}}\Phi_G(\pmb{R_G};t)
\end{equation}
The approximation involved in Eq.(\ref{eq:nabla_r1Phi}) results from the neglect of all angular derivations due to their occurrence
with coefficients decreasing faster than $R^{-1}$.
We proceed now to a quasiclassical approximation for the description of the center of mass translational motion, 
with two implications:

(i) $\pmb{R_G}\simeq v\pmb{u_y} t$ has a corresponding wavevector $\pmb{K_G}\simeq m v\pmb{u_y}$ and
the application of momentum operator $-i\nabla_{\pmb{R_G}}$
to $\Phi_G(\pmb{R_G})$ simply result into $m v \Phi_G(\pmb{R_G})$.
When this is done at the level of Eq(\ref{eq:j1}) one gets:
\begin{equation}\label{eq:j2}
\pmb{j}(t)=\frac{1}{m}\int d\pmb{r_2} \frac{1}{|\pmb{r_1}-\pmb{r_2}|^2} \mathrm{Im}\left[\Phi^*(\pmb{R};t) \frac{\partial}{\partial R} \Phi(\pmb{R};t)+imv|\Phi(\pmb{R};t)|^2\right]|\Phi_G(\pmb{R_G};t)|^2.
\end{equation}
(ii) No wavepacket spreading is allowed for $\Phi_G(\pmb{R_G};t)$ which is localized with an envelope
behaving as a $\delta$-like function, {\it i.e.}:
\begin{equation}\label{eq:PhiG}
|\Phi_G(\pmb{R_G})|^2\simeq \delta(R_G-vt).
\end{equation}
The integration over $\pmb{r_2}$ (with ${\mathrm d}\pmb{r_2}=2{\mathrm d}\pmb{R_G}$) finally leads to:
\begin{equation}\label{eq:j(t)}
\pmb{j}(t)=\frac{1}{m R^2}\mathrm{Im} \left.\left[\Phi^*(\pmb{R};t)\frac{\partial}{\partial R}\Phi(\pmb{R};t)\pmb{u_R} +imv|\Phi(\pmb{R};t)|^2 \pmb{u_y}\right]\right|_{\pmb{R}=2\pmb{r_1}-2vt\pmb{u_y}},
\end{equation}
with a rather intuitive interpretation of the two components of the flux.
The first {\i.e.} $\Phi^*(\pmb{R};t)\frac{\partial}{\partial R}\Phi(\pmb{R};t)\pmb{u_R}$ is merely the current density
generated by the expanding wavepacket in the center of mass frame, whereas the second corresponds
to the current associated with a density $\frac{|\Phi|^2}{R^2}$ travelling with a velocity $v$ along $\pmb{u_y}$.
The calculation can be further conducted analytically by deriving an asymptotic 
({\it i.e.} $R\to \infty$, $t \to\infty$) expression for $\Phi$ \cite{126}.
This is done using a time evolution expression
involving the Fourier transform Eq.(\ref{eq:FFT}). Actually, one has for large $R$, where the potentials can be considered as constant
and after the laser is turned off:
\begin{equation}\label{eq:evPhi}
\hat{\Phi}(k,\theta,\phi;t)=e^{-ik^2t/m}\hat{\Phi}(k,\theta,\phi)
\end{equation}
the solution being induced only by the radial part of the kinetic energy.
Returning back to the wavepacket in the coordinate space:
\begin{equation}\label{eq:evPhiFFT}
\Phi(R,\theta,\phi;t)=\frac{1}{\sqrt{2\pi}}\int_{-\infty}^{\infty}{\mathrm d}k\hat{\Phi}(k,\theta,\phi)e^{-ik^2t/m}e^{ikR}
\end{equation}
and replacing $\hat{\Phi}$ by its expression Eq.(\ref{eq:FFT}), one gets:
\begin{equation}\label{eq:evPhiFFT2}
\Phi(R,\theta,\phi;t)=\left(\frac{m}{4i\pi t}\right)^{1/2}\int_{0}^{\infty}{\mathrm d}R' \Phi(R',\theta,\phi)e^{im(R-R')^2/4 t}.
\end{equation}
Expanding the $R$-dependent part of the exponential as:
\begin{equation}\label{eq:exp1}
e^{\frac{i m}{4t}(R-R')^2}=e^{\frac{i m R^2}{4 t}}e^{-\frac{i m RR'}{2 t}} 
\left[ 1+ \left( e^{\frac{i m {R'}^2}{4t}} -1\right)\right]
\end{equation}
and observing \cite{22} that for large $t$:
\begin{equation}\label{eq:lim_exp}
\lim_{t\to \infty}\left|\left( e^{\frac{i m {R'}^2}{4 t}} -1\right)\right|=0,
\end{equation}
an asymptotic expression is obtained for $\Phi(R,\theta,\phi;t)$:
\begin{equation}\label{eq:Phi_as}
\Phi(R,\theta,\phi;t) \sim_{t\to \infty} 
\left( \frac{m}{2 it}\right)^{1/2}e^{\frac{i m R^2}{4t}}\hat{\Phi}\left(\frac{mR}{2t},\theta,\phi\right).
\end{equation}
While recasting Eq.(\ref{eq:Phi_as})
into Eq.(\ref{eq:j(t)}), a rather simple expression results for the asymptotic flux:
\begin{equation}\label{eq:j(t)_fin}
\pmb{j}(t)=\frac{m}{2Rt^2} \left.\left[\pmb{u_R} +\frac{2vt}{R}\pmb{u_y}\right] 
\left|\hat{\Phi}\left(\frac{mR}{2t},\theta,\phi\right)\right|^2  \right|_{\pmb{R}=2(\pmb{r_1}-vt\pmb{u_y})}
\end{equation}

The calculation of the projection of $\pmb{j}$ on $\pmb{u_y}$ (cf Eq.\ref{eq:DN}) requires the vector relation of Eq.(\ref{eq:A7})
that finally leads to:
\begin{equation}\label{eq:j_uy}
\pmb{j}\cdot\pmb{u_y}=\frac{mD}{4 t^2}\frac{1}{[\rho^2+(D-vt)^2]^2} \left|\hat\Phi\left(\frac{mR}{2t},\theta,\phi\right)\right|^2.
\end{equation}
where $R$ depends on $t$ as given by Eq.(\ref{eq:A3}).

The final step is to transform the time integration of $\pmb{j}\cdot \pmb{u_y}$, involved in Eq.(\ref{eq:DN}), into an integration over the kinetic momentum.
We proceed to a change of variable:
\begin{equation}\label{eq:k}
k=[\rho^2+(D-vt)^2]^{1/2}\frac{mv}{D}=\frac{R}{2}\frac{mv}{D},
\end{equation}
the physical meaning of which will be clarified hereafter. Straightforward calculations show that Eq.(\ref{eq:k}) can be inverted as:
\begin{equation}\label{eq:t_mp}
t=
\left \{ \begin{array}{l}
\frac{D}{v}\left(1-\frac{(k^2-k_\rho^2)^{1/2}}{mv}\right)\ \  \mathrm {for}\  t \in \left[ 0,\frac{D}{v}\right ]\ \ 
(k \in [k_\rho,(k_\rho^2+m^2v^2)^{1/2}])\\
\frac{D}{v}\left(1+\frac{(k^2-k_\rho^2)^{1/2}}{mv}\right)\ \  \mathrm {for}\  t \in \left[ \frac{D}{v},+\infty \right ]\ \ 
(k \in [k_\rho,+\infty])\\
\end{array} \right.,
\end{equation}
upon the introduction of the notation
\begin{equation}\label{eq:krho_int}
k_{\rho}=mv\frac{\rho}{D}
\end{equation}
and leads to:
\begin{equation}\label{eq:dt}
{\mathrm d}t=\mp\frac{D}{mv^2}\frac{k}{\left( k^2-k_{\rho}^2 \right)^{1/2}}{\mathrm d}k.
\end{equation}
The $\mp$ signs correspond to the two time intervals depicted in Eqs(\ref{eq:t_mp}).
The time dependent argument of
$\hat\Phi$ in Eq.(\ref{eq:Phi_as}) can than be expressed using the two variables $k$ (Eq.(\ref{eq:k})) and $k_\rho$ (Eq.(\ref{eq:krho_int})) as:
\begin{equation}\label{eq:k_Phi}
\frac{mR}{2t}=\frac{m}{t}[\rho^2+(D-vt)^2]^{1/2}=\frac{D}{vt}k=k\left(1 \mp \frac{(k^2-k_\rho^2)^{1/2}}{mv}\right)^{-1}.
\end{equation}
The experimental conditions, are such that the velocity $v$ of the molecular beam is much greater than
the fragments relative velocity. We can thus consider
$(k^2-k_\rho^2)^{1/2}/mv$ as negligible when compared to 1, taking into account that $D$ is much larger than $\rho$.
The resulting approximation, namely:
\begin{equation}\label{eq:t_app}
t\simeq \frac{D}{v}\ \ \ \ {\mathrm and} \ \ \ \ \frac{mR}{2t}\simeq k
\end{equation}
merely means that the time needed for a fragment to reach the pixel $M(\rho, \alpha)$
is approximately the same as the one needed for the center of mass $G$ to reach the center $O'$ of the detector.
In the framework of this approximation, the meaning of $k_\rho=m \rho v/D \simeq m\rho/t$ (defined by Eq.(\ref{eq:krho_int}) ) 
is also clear: {\it i.e.} the projection of the kinetic momentum $k$ on the detector plane.
Finaly we obtain for the time integrated flux:
\begin{subequations}\label{eq:int_j_uy_dt}
\begin{equation}\label{eq:int_j_uy_dt_a}
\int_0^{\infty}\pmb{j}\cdot\pmb{u_y} {\mathrm d}t= \frac{m^2v^2}{4 D^2 } \left[\int_{k\rho}^{\infty} + 
\int_{k\rho}^{(k_\rho^2+m^2v^2)^{1/2}}\frac{\left|\hat\Phi\left(k,\theta,\phi\right)\right|^2}{k\left( k^2-k_{\rho}^2 \right)^{1/2}}{\mathrm d}k \right]
\end{equation}
\begin{equation}\label{eq:int_j_uy_dt_b}
= \frac{m^2v^2}{2 D^2 }\int_{k\rho}^{\infty}\frac{\left|\hat\Phi\left(k,\theta,\phi\right)\right|^2}{k\left( k^2-k_{\rho}^2 \right)^{1/2}}  {\mathrm d}k.
\end{equation}
\end{subequations}
where the upper bond of the second integral in Eq.(\ref{eq:int_j_uy_dt_a}) has been extended up to $+\infty$
considering that $|\hat\Phi(k,\theta,\phi)|=0$ for $k > mv$,
which is equivalent to state that
the center of mass kinetic momentum $2mv$ is much larger than the relative momentum of photofragments $k$.
Recasting Eqs(\ref{eq:int_j_uy_dt}) in Eq.(\ref{eq:DN}), taking into account cylindrical symmetry over $\phi$ and
calculating the pre-integral factor as:
\begin{equation}\label{eq:abel_fact}
\frac{m^2v^2}{D^2}{\mathrm d}S=\frac{mv\rho}{D}\frac{mv {\mathrm d}\rho}{D}{\mathrm d}\alpha=k_\rho {\mathrm d} k_\rho {\mathrm d}\alpha.
\end{equation}
one finally gets:
\begin{equation}\label{eq:DN_1}
{\mathrm d}N(k_\rho,\alpha)= N k_\rho {\mathrm d}k_\rho {\mathrm d}\alpha  \frac{1}{2} \int_{k_\rho}^{\infty}  \frac{\left|\hat\Phi\left(k,\theta\right)\right|^2}{k\left( k^2-k_{\rho}^2 \right)^{1/2}}  {\mathrm d}k.
\end{equation}
The dependence over $\theta$ of the right-hand-side of Eq.(\ref{eq:DN_1})
has to be expressed in terms of $\alpha$, referring to the frame transformation Eq.(\ref{eq:A9})
\begin{equation}\label{eq:cos_th}
\cos\theta= \frac{k_\rho}{k}\cos\alpha
\end{equation}
in such a way that, ultimately ${\mathrm d} N$ is written
only in terms of the variables $k_\rho$ and $\alpha$,
with the parameters $v$ and $D$ characterizing the experimental setup:
\begin{equation}\label{eq:DN_2}
{\mathrm d}N(k_\rho,\alpha)= N k_\rho {\mathrm d}k_\rho {\mathrm d}\alpha  \frac{1}{2} \int_{k_\rho}^{\infty}  \frac{\left|\hat\Phi\left(k,\arccos(k_\rho/k \cos\alpha)\right)\right|^2}{k\left( k^2-k_{\rho}^2 \right)^{1/2}}  {\mathrm d}k.
\end{equation}
The probability to record a hydrogen atom on the surface element $\mathrm{d}S$ (pixel $M$) located at $\rho$, $\alpha$ on the MCD (with a kinetic momentum $k_\rho$) is obtained by a proper
normalization:
\begin{equation}\label{eq:DN_3}
P(k_\rho,\alpha) {\mathrm d}S = \frac{1}{N}{\mathrm d}N(k_\rho,\alpha)=\frac{{\mathrm d}S}{2} \int_{k_\rho}^{\infty}  \frac{\left|\hat\Phi\left(k,\arccos(k_\rho/k \cos\alpha)\right)\right|^2}{k\left( k^2-k_{\rho}^2 \right)^{1/2}}  {\mathrm d}k.
\end{equation}
It is interesting to note that the two probabilities ${\cal P}(k,\theta)$ (given in Eq.\ref{eq:Pkth}) and $P(k_\rho,\alpha)$ (Eq.(\ref{eq:DN_3}))
are simply connected by:
\begin{equation}\label{eq:DN_4}
P(k_\rho,\alpha)= \frac{1}{2} \int_{k_\rho}^{\infty}  \frac{{\cal P}(k,\theta)}{k\left( k^2-k_{\rho}^2 \right)^{1/2}}  {\mathrm d}k.
\end{equation}

Two remarks are in order:

i) both equations Eq.(\ref{eq:DN_3}) and Eq.(\ref{eq:DN_4}) involve a singularity at $k=k_\rho$.
This difficulty can be overcame by a partial integration leading to:
\begin{equation}\label{eq:abel_2}
\begin{split}
P(k_\rho,\alpha)=\frac{1}{2 k_{\rho}}\arccos\left(\frac{k_{\rho}}{k} \right) {\cal P}\left(k,\theta\right)
\Big|_{k=k{\rho}}^{k=\infty}
-\frac{1}{2 k_\rho}\int_{k_{\rho}}^{\infty}\arccos\left(\frac{k_{\rho}}{k} \right) \frac{{\mathrm d} }{{\mathrm d} k}{\cal P}(k,\theta) {\mathrm d}k
\end{split}
\end{equation}
The integrated term in the right-hand-side of Eq.(\ref{eq:abel_2}) is null, due to the fact that ${\cal P}(k,\theta)\Big|_{k=\infty}=0$.
As for the total derivative with respect to $k$ of ${\cal P}(k,\theta)$, it results into:
\begin{equation}\label{eq:abel_2_res}
\begin{split}
\frac{{\mathrm d} }{{\mathrm d} k}{\cal P}(k,\theta)=
\frac{\partial {\cal P}}{\partial k}+
\frac{k_{\rho}\cos\alpha}{k(k^2-k_{\rho}^2\cos^2\alpha)^{1/2}}\frac{\partial {\cal P}}{\partial \theta}.
\end{split}
\end{equation}
When recasting Eq.(\ref{eq:abel_2_res}) into Eq.(\ref{eq:abel_2}) one obtains:
\begin{equation}\label{eq:abel_3}
\begin{split}
P(k_\rho,\alpha)=
-\frac{1}{2 k_\rho}\int_{k_{\rho}}^{\infty}\arccos\left(\frac{k_{\rho}}{k} \right) \left[  \frac{\partial {\cal P}}{\partial k}+
\frac{k_{\rho}\cos\alpha}{k(k^2-k_{\rho}^2\cos^2\alpha)^{1/2}}\frac{\partial {\cal P}}{\partial \theta}\right]{\mathrm d}k
\end{split}
\end{equation}
For $k \simeq k_\rho$, the singularity in the coefficient of $\frac{\partial{\cal P}}{\partial \theta}$ may only occur for $\alpha=0$.
This is fortunately compensated by the $\arccos(k_\rho/k)$ term of Eq.(\ref{eq:abel_2}). Actually expanding Eq.(\ref{eq:abel_2}) in terms of powers of $(1-k_\rho/k)$
one ends up with a non-singular behavior, {\it i.e.} $\frac{1}{k_\rho}\frac{\partial {\cal P}}{\partial \theta}$ for the integrand in the vicinity of $k_\rho$.

ii) Despite the fact that the experimental situation we are describing is not amenable to a simple mapping
on the detector plate of a photodissociation that had already occured in the center of mass frame (as a figure \ref{fig:shexp_1}),
it turns out that Eq.(\ref{eq:DN_3}) can finally be recast in terms of the commonly used Abel transformation \cite{225}:
\begin{equation}\label{eq:Abel_P}
P(k_\rho,\alpha)=\frac{1}{4}{\cal A}\left[ \left|\frac{\hat\Phi\left(k,\arccos(k_\rho/k \cos\alpha)\right)}{k}\right|^2\right],
\end{equation}
where $\cal A$ is defined as \cite{225}:
\begin{equation}\label{eq:Abel_def}
{\cal A}[f(k)](x)=2 \int_{x}^{\infty}  \frac{f(k)k}{\left( k^2-x^2 \right)^{1/2}}  {\mathrm d}k.
\end{equation}
In connection with this, it is worthwhile considering the full Fourier transform of the wavefunction describing
the relative motion (in contrast to the one carried in Eq.\ref{eq:FFT}):
\begin{equation}\label{eq:Psi_3k}
\hat \Psi(\pmb{k})=\frac{1}{(2\pi)^{3/2}}\int\!\!\!\int\!\!\!\int {\mathrm d} \pmb{R} \Psi(\pmb{R}) e^{-i \pmb{k}\pmb{R}}
\end{equation}
It can be shown by following the same derivations as Eqs(\ref{eq:evPhi}-\ref{eq:Phi_as}) \cite{126,22}, 
that asymptotically ({\it i.e.} for $t \to +\infty$ and $R \to +\infty$), one has:

\begin{equation}\label{eq:PsiPhi}
\Psi(\pmb{R};t)=\left(\frac{m}{2 i t}\right)^{3/2}
e^{-imR^2/4t}\hat{\Psi}\left(\frac{m \pmb{R}}{2 t}\right)
\end{equation}
which, combined with Eqs(\ref{eq:DefPhi} and \ref{eq:Phi_as}), implies that:
\begin{equation}\label{eq:ModPsiPhi}
|\hat{\Psi}(\pmb{k};t)|^2=\left|\frac{\hat{\Phi}(\pmb{k};t)}{k}\right|^2,
\end{equation}
The probability in Eq.(\ref{eq:Abel_P}) appears now as
the Abel transform of $\left|\hat\Psi\left(k,\arccos(k_\rho/k \cos\alpha)\right)\right|^2$.

\subsection{Rovibrationally averaged spectra}

The probability distributions which are calculated in the previous paragraph refer to a given initial state
($g$, $v$, $N$, $M_N$) involved in the determination of $\Phi(t=0)$ through Eq.(\ref{eq:Phi0_chi})
such that the quantity resulting from Eq.(\ref{eq:Abel_P}) is actually $P_{v,N,M_N}(k_\rho,\alpha)$ using a full notation.
An averaging over the initial ro-vibrational populations is thus required to reach the experimental spectra.
As the rotational states $N$ are $(2N+1)$ times degenerated, a summation can be carried out over $M_N$,
leading to:
\begin{equation}\label{eq:pop_1}
P_{v,N}(k_\rho,\alpha)=\frac{1}{2N+1}\sum_{M_N=0}^N c_{M_N} P_{v,N,M_N}(k_\rho,\alpha),
\end{equation}
where $c_0=1$ and $c_{M_N}=2$ (for $M_N \ne 0$), due to the fact that the total Hamiltonian does not depend
upon the sign of $M_N$. 
The homonuclear diatomic character of H$_2^+$ implies a total wavefunction (accounting for the nuclear spin) that is antisymmetric with respect
to the interchange of identical nuclei. To ensure such a property the total nuclear spin number $T$ must be either $0$ (associated with even $N$),
or $1$ (associated with odd $N$). Due to very rare singlet ($T=0$) - triplet ($T=1$) transitions, molecular hydrogen mainly consists
of two distinct species: parahydrogen ($T=0$) and orthohydrogen ($T=1$). The occurrence of ortho states is three times more probable
than the para ones \cite{14}. This nuclear spin statistics is accounted for, by a weighting coefficient $g_N$, affecting the rotational
populations $N$, such that
\begin{equation}\label{eq:pop_2}
g_N=
\left\{ \begin{array}{c}
1 \textrm{ for even }N\\
3 \textrm{ for odd } N
\end{array}
\right.
\end{equation}
Apart from this factor, rotational populations are also thermally weighted, according to a Boltzman distribution described by a rotational temperature
$T_v$ depending on the initial vibrational level. The weighting coefficient is given by:
\begin{equation}\label{eq:pop_3}
b_N=\exp\left[-\frac{\Delta E(v,N)}{k_\beta T_v}\right].
\end{equation}
where $k_\beta$ stands for the Boltzman constant and $\Delta E(v,N)$ for the rotational energies resulting from the solution of Eq.(\ref{eq:radSE}):
\begin{equation}\label{eq:pop_4}
\Delta E(v,N)=E(v,N)-E(v,0).
\end{equation}
The rotationally averaged probabilities resulting from these considerations are
\begin{equation}\label{eq:pop_5}
P_v(k_\rho,\alpha)=\frac{1}{Q_v}\sum_{N}b_N g_N P_{v,N}(k_\rho,\alpha),
\end{equation}
where
$Q_v$ is a normalization factor:
\begin{equation}\label{eq:pop_6}
Q_v=\sum_{N}b_N g_N.
\end{equation}
The comparison with experimental spectra has also to take into account initial vibrational populations $a(v)$
({\it i.e.} as they result from the electric discharge acting over H$_2$, prior to the laser excitation):
\begin{equation}\label{eq:pop_7}
P(k_\rho,\alpha)=\frac{1}{Q}\sum_{v}a(v) P_{v}(k_\rho,\alpha),
\end{equation}
with
\begin{equation}\label{eq:pop_8}
Q=\sum_{v}a(v).
\end{equation}
We note that the informations concerning the initial vibrational distribution $a(v)$ as well as the corresponding rotational temperature $T_v$,
have to be provided by experimental measurements.

\subsection{Laser spatial intensity averaging}

Although particular attention is devoted in the experimental measurements for obtaining a well focussed ion beam, the
molecules are actually excited by different field amplitudes according to their positions ($x$, $y$), due to a non-homogeneous spatial intensity
distribution $I(x,y)$ in the laser beam. It turns out that an average over these non-homogeneities has a basic importance when attempting
a quantitative interpretation of experimental data, as will be clear in the next section. The average implies a double spatial integration
over the variables $x$ and $y$ (see figure \ref{fig:int_beam}):
\begin{equation}\label{eq:int_1}
P(k_\rho,\alpha)=\int_{-L}^{L} dx \int_{-\infty}^{+\infty}  P(k_\rho,\alpha;I(x,y)) dy.
\end{equation}
$x$ being limited to a finite interval $2L$ measuring the width of the ion beam. The integrand itself is nothing but the probability
calculated in Eq.(\ref{eq:pop_7}) 
with as an additional argument the intensity  $I(x,y)$, for which
this probability is calculated, {\it i.e.} $P(k_\rho,\alpha;I(x,y))$. For parity reasons Eq.(\ref{eq:int_1})
may be also written as:
\begin{equation}\label{eq:int_2}
P(k_\rho,\alpha)=4\int_{0}^{L} dx \int_{0}^{+\infty}  P(k_\rho,\alpha;I(x,y)) dy.
\end{equation}
A gaussian shape is assumed for the 2D behavior of the laser within its focus area:
\begin{equation}\label{eq:int_3}
I(x,y)=I_0\exp\left[-\frac{x^2}{r_x^2}\right]\exp\left[-\frac{y^2}{r_y^2}\right]
\end{equation}
where $r_x$ and $r_y$ are the radii of the focal area, such that $I(r_x,r_y)=I_0/e^2$. These parameters are obtained 
from the experimental setup \cite{12} as:
\begin{equation}\label{eq:int_4}
r_{x,y}=\frac{\lambda f}{2 \pi b_{x,y}}
\end{equation}
where $\lambda$ is the laser wavelength and $f$ is the focal length of the parabolic mirror focusing the laser beam.
$b_x$ and $b_y$ correspond to the extensions in each direction $x$ and $y$ where 50\% of the energy is dissipated.
The peak intensity is calculated whence the total pulse energy $E_0$ and an autocorrelation time $t_{ac}$ have been measured:
\begin{equation}\label{eq:int_9}
I_0=\frac{2 \sqrt{ 2 \ln 2} E_0}{\pi^{3/2}r_x r_y t_{ac}}.
\end{equation}
The double integration involved in Eq.(\ref{eq:int_2}) can be conducted, by integrating first over $y$, referring to a variable change:
\begin{subequations}\label{eq:int_12}
\begin{equation}\label{eq:int_12a}
y=r_y\left[-\ln\left(\frac{I}{I_x}\right)\right]^{1/2}
\end{equation}
\begin{equation}\label{eq:int_12b}
dy=-\frac{r_y}{2}\left[-\ln\left(\frac{I}{I_x}\right)\right]^{-1/2}\frac{dI}{I},
\end{equation}
\end{subequations}
where $I_x=I_0\exp\left[-(x/r_x)^2\right]$. The result is:
\begin{equation}\label{eq:int_13}
P(k_\rho,\alpha;I_x)=2 r_y \int_{0}^{I_x}  \frac{ P(k_\rho,\alpha;I)}{I \left[-\ln\left(\frac{I}{I_x}\right)\right]^{1/2}} dI.
\end{equation}
Two singularities affect such an expression; namely at $I=0$ and at $I=I_x$. The first has no consequence, as for $I=0$, 
$P(k_\rho,\alpha;I=0)=0$.
The second can be avoided when integrating by part:
\begin{equation}\label{eq:int_14}
P(k_\rho,\alpha;I_x)=2 r_y \left[ - 2\sqrt{I/I_x} P(k_\rho,\alpha;I) + \int_{0}^{I_x} 2\sqrt{I/I_x}\frac{d}{dI}P(k_\rho,\alpha;I)dI  \right]
\end{equation}
An identical procedure is then applied for the second integral over $x$. The final result is
\begin{equation}\label{eq:int_15}
P(k_\rho,\alpha)=2 r_x \left[ \left. - 2\sqrt{I_x/I_0} P(k_\rho,\alpha;I_x) \right|^{I_x=I_0}_{I_x=I_L} + \int_{I_L}^{I_0} 2\sqrt{I_x/I_0}\frac{d}{dI_x}P(k_\rho,\alpha;I_x)dI_x  \right]
\end{equation}
where $I_L$ is the field intensity at ($x=L$, $y=0$). 

\pagebreak

\section{Results}
\label{sec:Results}
This section presents the results of the simulation and interpretation of experimental data.
Among the experimental results obtained in H.Figger's group \cite{12,20,21} three
are retained. 
The laser intensities $I_0$ have been slightly adjusted with respect to laboratory measurements of the total pulse
energy $E_0$ and autocorrelation time $t_{ac}$, to fit experimental spectra. The resulting
parameters are collected in Table \ref{tab:exp_params}. 
The laser pulse carrier wavelength is
$\lambda$=785 nm.
\vspace{0.5cm}
\begin{table} [!htp]
\begin{center}
\begin{tabular}{|c|c|c|c|}
\hline
E$_0$ (mJ) & t$_{ac}$ (fs) & I$_0$ (TW/cm$^2$)\\
\hline
0.3 & 228 & 7.5 \\
\hline
0.5 & 234 & 9.5 \\
\hline
0.7 & 240 & 16.0\\
\hline
\end{tabular}
\end{center}
\caption{\footnotesize \footnotesize Total pulse energy E$_0$ and autocorrelation time t$_{ac}$ of the laser field. $I_0$ is the maximal
field intensity value.}\label{tab:exp_params}
\end{table}
\vspace{0.5cm}
In the calculations the intensity spatio-temporal distribution is assumed to be:
\begin{equation}\label{eq:resint_1}
I(x,y;t)=I_0\exp\left[-\frac{x^2}{r_x^2}\right]\exp\left[-\frac{y^2}{r_y^2}\right]\exp\left[-\frac{2t^2}{w_t^2}\right]
\end{equation}
with the relations between the parameters as given by Eqs(\ref{eq:int_4}, \ref{eq:int_9}).
The widht of the molecular jet is $L$=50 $\mu$m, its velocity is $v$=10$^6$ m/s
and focal area parameters values are $b_x$=2.6 mm, $b_y$=2.4 mm, $f$=1000 mm,
resulting into $r_x$=48$\mu$m, $r_y$=52$\mu$m.
In the calculations described below the parameter $w_t=t_{ac}/2\sqrt{ln 2}$, which define the laser pulse temporal shape,
have been taken equal to 140 fs.
\begin{figure}[!htp]
\begin{center}
   \includegraphics[scale=0.5,angle=0]{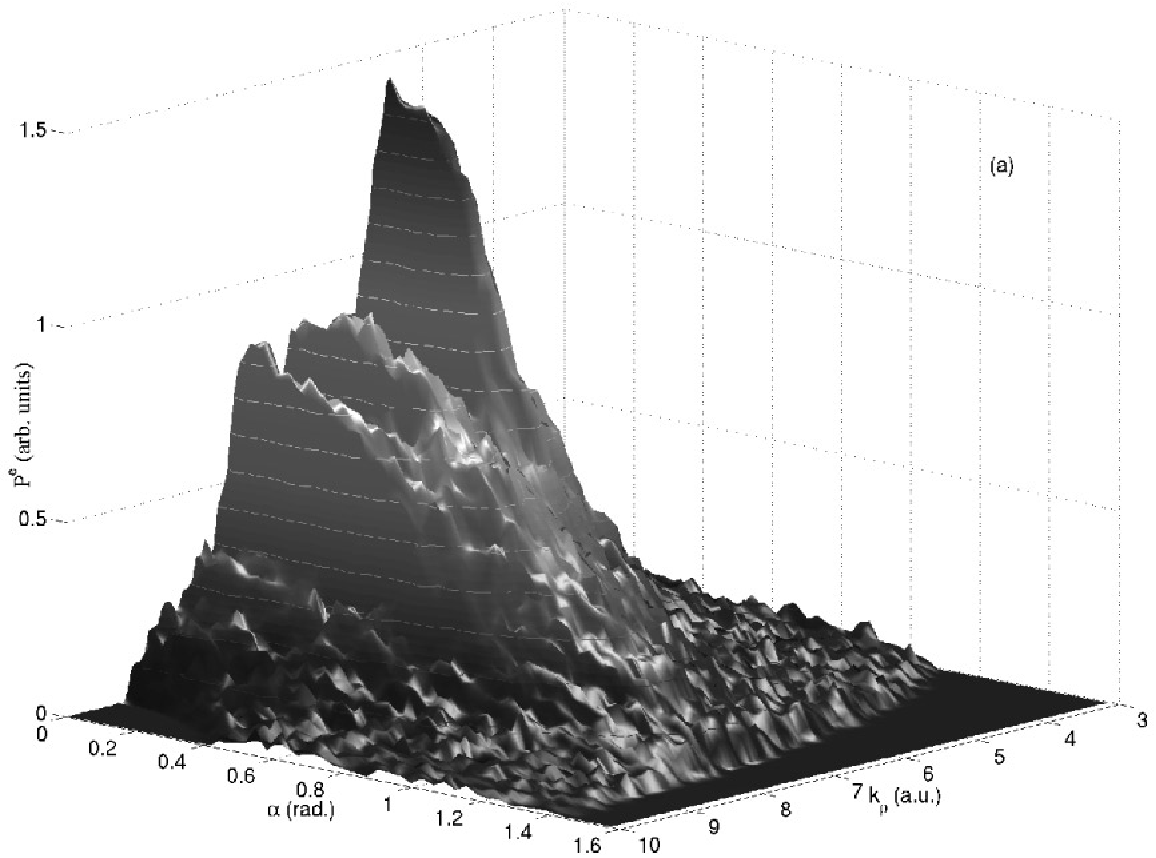}\\
   \includegraphics[scale=0.5,angle=0]{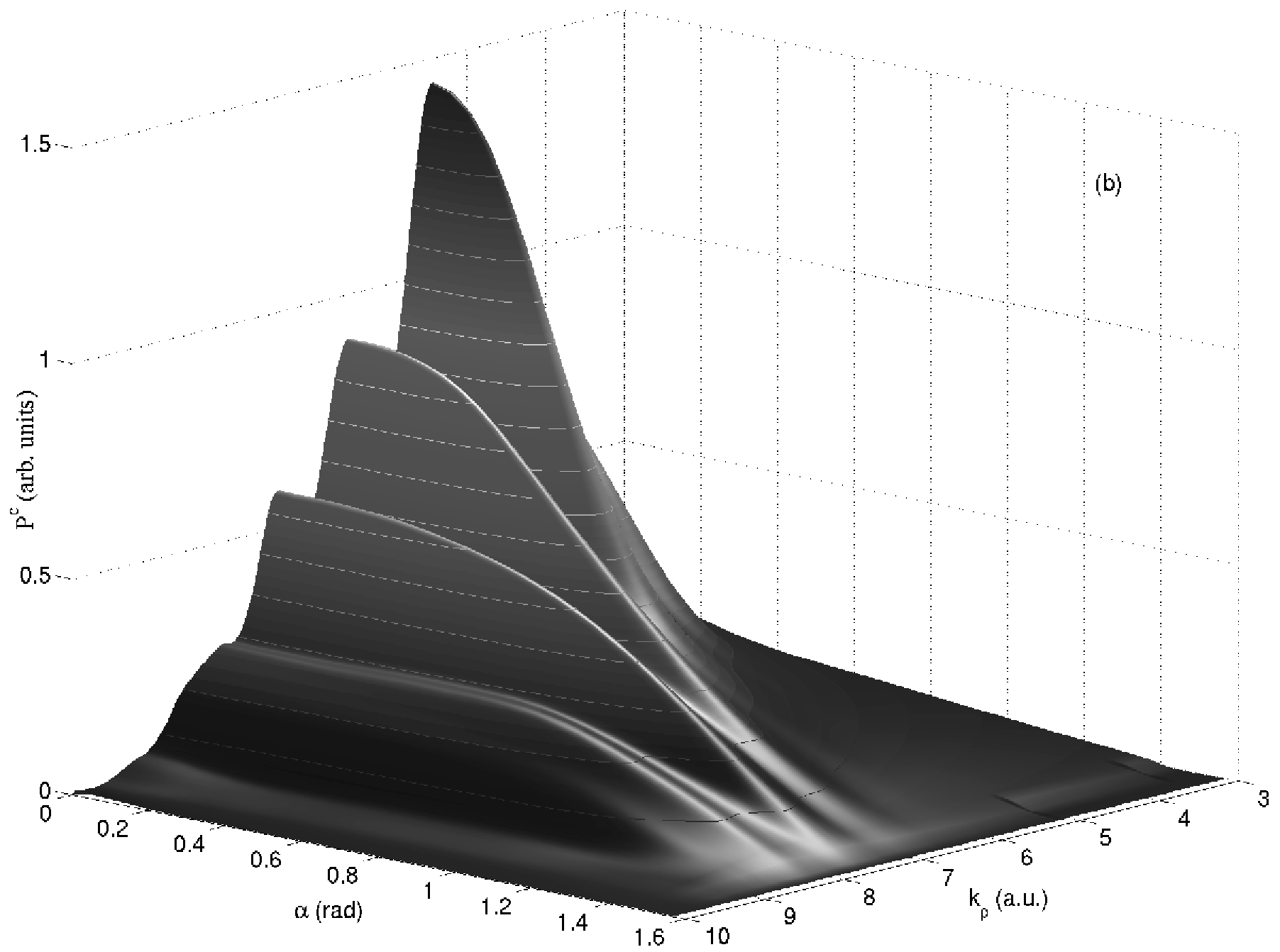}\\
\end{center}
\end{figure}
\begin{figure}[!ht]
\begin{center}
   \includegraphics[scale=0.5,angle=0]{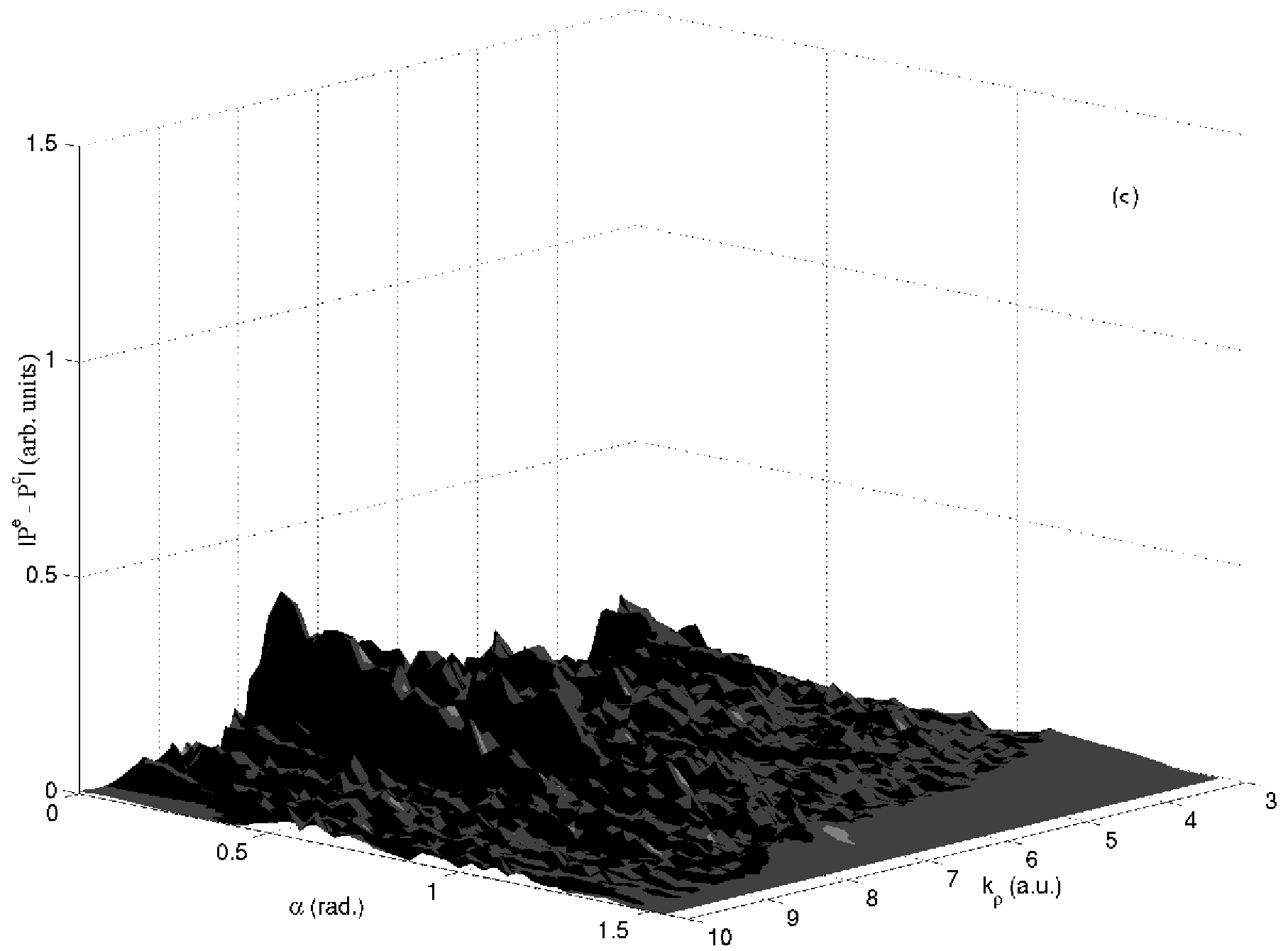}
\end{center}
\caption{\footnotesize (a) - The three dimensional representation of the experimental result corresponding to
E$_0$=0.7 mJ (see the third line in Tab.(\ref{tab:exp_params});
(b) - The corresponding calculation result; (c) - The difference between experimental and calculated spectra.}\label{fig:3D_res}
\end{figure}

\begin{figure}[!ht]
\begin{center}
   \includegraphics[scale=0.6,angle=270]{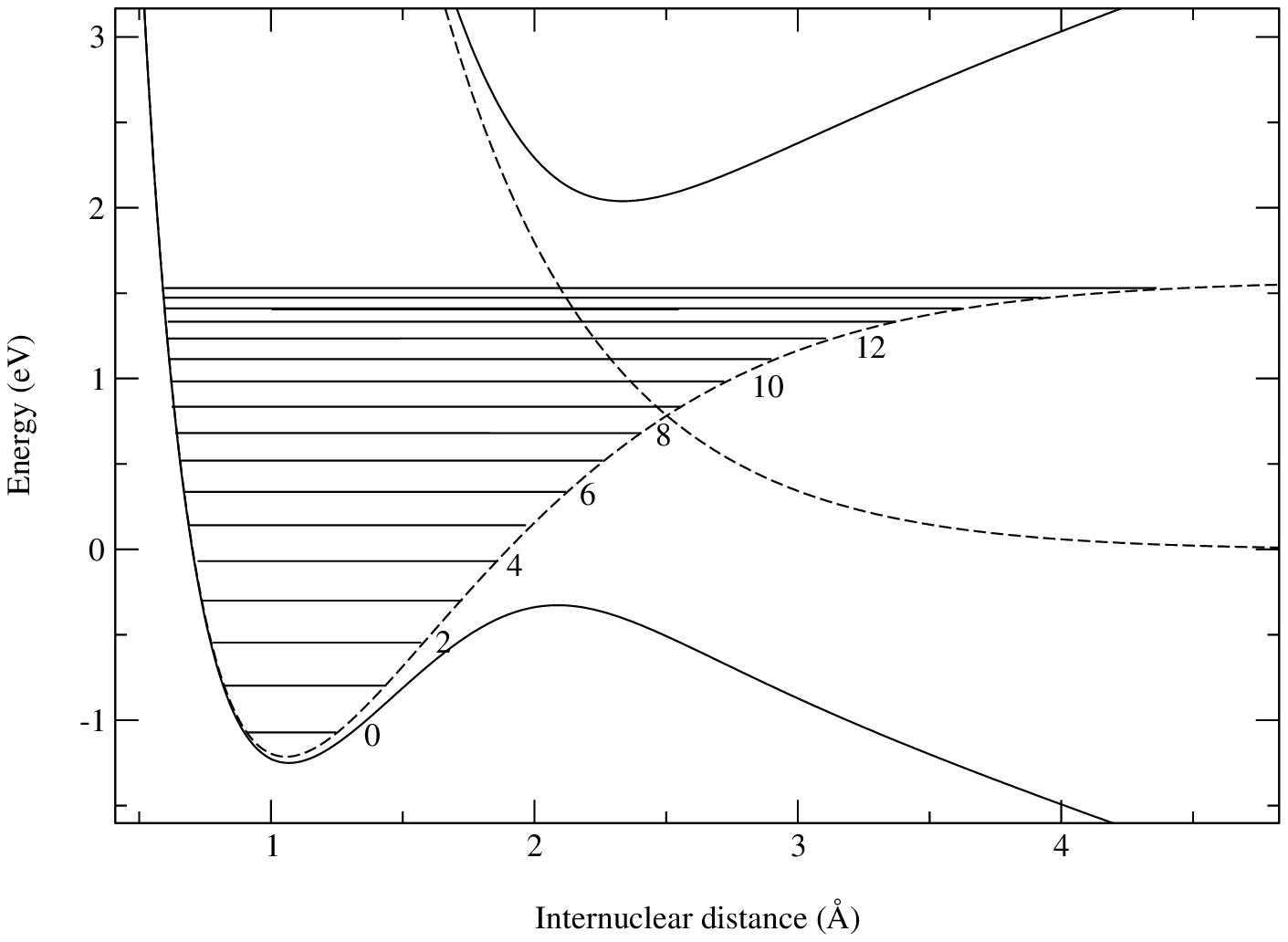}
\end{center}
\caption{\footnotesize 2D view of the dressed adiabatic potential curves of H$_2^+$ (solid lines)
(with a continuous wave laser of wavelength $\lambda$=785 nm
and intensity $I_0$=1.6$\times$10$^{13}$ W/cm$^2$) and the corresponding diabatic BO electronic states (dashed lines). Are also indicated the
 eigenvalues of the field-free vibrational levels.}\label{fig:ptls_2d}
\end{figure}

Figure \ref{fig:3D_res} displays three-dimensional representations of the dissociation probabilities as a function of
their angular ($\alpha$) and kinetic ($k_\rho$) distributions. The upper diagram corresponds to the experimental results
$P^e$ \cite{20} for the laser excitation parameters indicated on the last row of Table \ref{tab:exp_params}.
The lowest two diagrams give the calculated spectrum ($P^c$) and the absolute value of the difference $\left|P^e-P^c\right|$, for
the same laser characteristics, at the same scale.
The normalization is such that:
\begin{equation}
\int_0^{\infty}k_\rho dk_\rho \int_0^{\pi/2} P^{e,c}(k_\rho,\alpha) d\alpha=1.
\end{equation}
The successive peaks that are obtained correspond to photofragments
arising from different vibrational levels $v$ of the parent ion H$_2^+$. The energies of the levels are positioned
in figure \ref{fig:ptls_2d} on the rotationless dressed molecular potentials resulting from the diagonalization
of the radiative interaction at fixed angle $\theta=0$.

The most important peak  (at $\alpha=0$) corresponds to $v$=7 and is followed in decreasing order by the peaks
assigned to $v$=8,9,10. The peak resulting from the dissociation of $v$=6, with a much smaller contribution is hidden by the peak $v$=7,
whereas those resulting from $v$=11,12 are in the blue-wing of $v$=10. It is interesting to note that, on the lowest diagram, the largest error affects
the peak resulting from v=9, all others being well represented. This is to be relationed with the particular energy of $v$=9
(see figure \ref{fig:ptls_2d}) very close to the avoided crossing of the dressed potentials. The characteristics of this
region being very sensitive to the laser spatial and temporal intensity distributions, even small deviations with respect to
experimental evaluations may lead to appreciable differences explaining figure \ref{fig:3D_res}(c).

\begin{figure}[!htp]
\begin{center}
   \includegraphics[scale=0.6,angle=0]{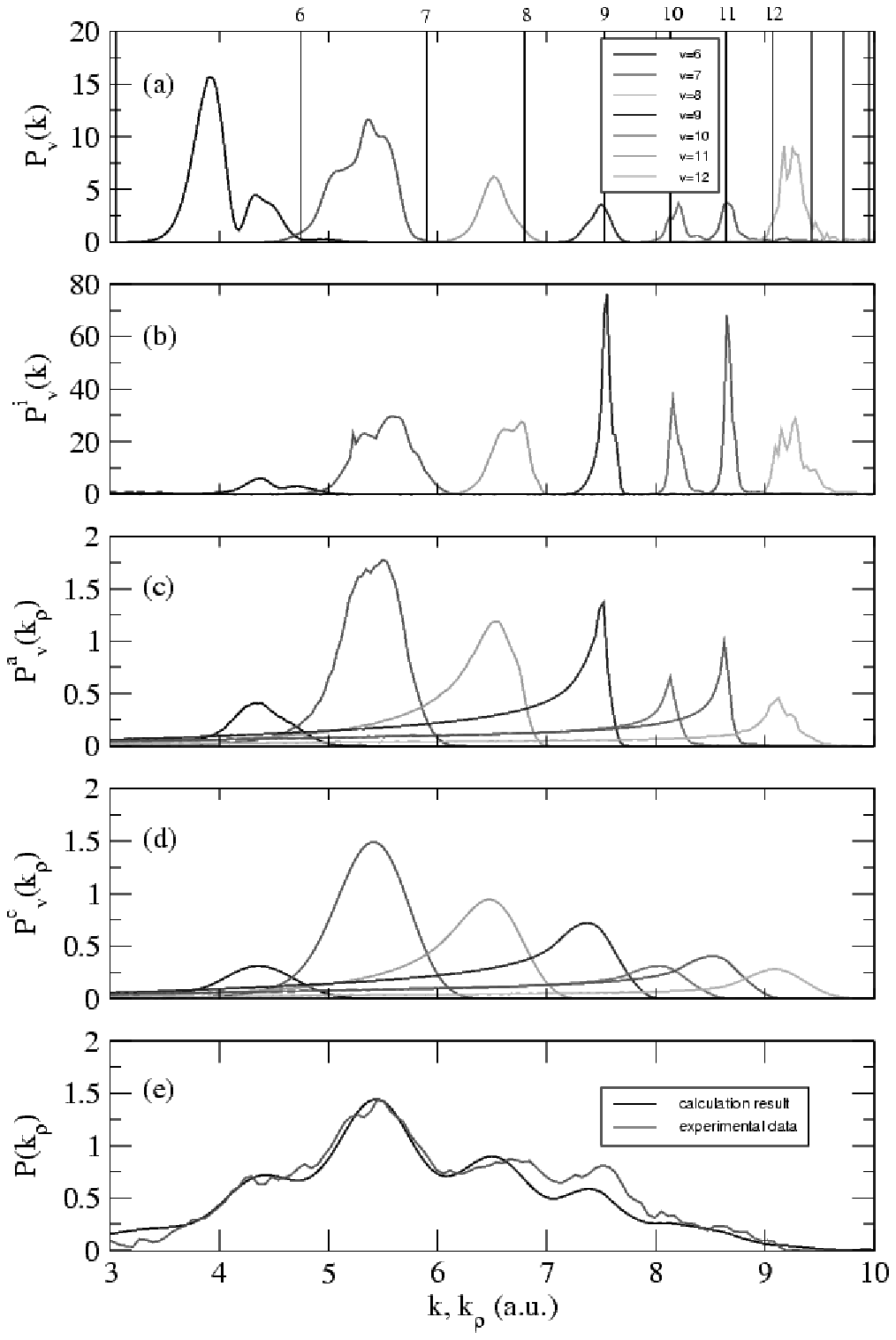}
\end{center}
\caption{\footnotesize
Successive steps for a theory-versus-experiment comparison of photodissociation spectrum for a laser energy 0.7 mJ
(last row in the Table \ref{tab:exp_params}). Panel (a) gives the individual probabilities of initial vibrational levels
$v$=6,...,12 in the molecular frame, for $\theta=0$, and as a function of $k$. Panel (b) takes into account the spatial intensity
distribution of the laser. Panel (c) displays the intensity averaged probabilities after Abel transformation bringing them
in the laboratory frame, for $\alpha=0$ and as a function of $k_\rho$. Panel (d), same as c but after convolution by the
detector resolution window. Panel (e) sums up all individual v contributions and compares with the experimental spectrum.
}\label{fig:intapop}
\end{figure}

In figure \ref{fig:intapop} we show the four main steps to obtain the photofragments distribution P$^c$, which may be compared with the experimental one.
On each step we plot the cut of the resulting distribution at $\theta=0$ for ${\cal P}(k,\theta)$, or $\alpha=0$ for $P(k_\rho,\alpha)$.
The upper panel gives the photodissociation probabilities starting from individual
vibrational levels of H$_2^+$, as calculated in the molecular frame for $\theta$=0 and as a function of $k$,
for laser characteristics corresponding to the last row of Table \ref{tab:exp_params}.
The vertical lines illustrate the theoretical energies of the vibrational levels of H$_2^+$ in a field-free situation.
As expected, from the examination of figure \ref{fig:ptls_2d}, the maxima of the peaks with $v<9$ are shifted
down and these of $v>9$ are shifted up, due to the
radiative coupling. As to the height of the successive peaks, a decrease from $v$=6 to $v$=9 is observed.
This, however, does not mean that $v$=9 is less dissociated than $v$=6,
as the information which is displayed concerns a cut at angle $\theta=0$.

The major effect, when attempting a theory-versus-experiment comparison, is the role played by the spatial intensity
distribution of the laser, that so far, has been neglected by referring to the large laser focal area with respect to the
diameter of the ion beam \cite{Kondorskiy}. Spatial averaging brings into interplay molecules interacting with radiative fields having
intensities less than the maximum value $I_0$. This may lead to very large effects on some vibrational levels as is seen
in figure \ref{fig:intapop} panel (b). More precisely, the peak associated with $v$=6 is nearly washed out, whereas those
describing photodissociation starting from $v$=9 and $v$=11 seem to be enhanced. Only the laser maximum intensity leads to
a barrier lowering (bond softening) mechanism for $v$=6. When a spatial average is carried out, with the inclusion of
lower intensities, the photodissociation from $v$=6 is severely inhibited due to very low tunneling, which explains
the flat behavior of ${\cal P}^i_v(k)$ for $v$=6.The vibrational states $v$=7,8 are also affected by this effect
but in a  lesser extend as they are closer to the top of the lower adiabatic potential barrier. The level $v$=9 is again
in a particular situation, energetically lying on the very top of the barrier. In other words, its photodissociation
is not inhibited by any spatial intensity averaging, and this is why it leads to a narrower and higher peak, than
the ones originating from $v$=7,8. The narrowing of the peak, in particular, is in relation with the fact that only a
limited energy range corresponds to efficient dissociation within the open gate between the lower and upper adiabatic potentials
that gets narrower with decreasing intensities. The behavior of $v$=11 deserves particular interest, as its photodissociation
is rather through a vibrational trapping mechanism involving the upper adiabatic potential. Lower the field intensity and lesser is
the efficiency of this trapping. The spatial intensity averaging of the laser leads as a consequence to better relative dissociation
from $v$=11 resulting into a high peak.

Figure \ref{fig:intapop} panel (c) displays the dissociation probabilities $P^a_v(k_\rho)$ as functions of $k_\rho$
after the Abel transformation Eq.(\ref{eq:DN_4}).
The basic difference with ${\cal P}^i_v(k)$ (panel b) is the rise of long range red tails of the individual peaks, especially for $v\ge 9$.
This is due to the nonlinear features of the Abel transformation, mixing up a whole range of $\theta$-dependent
probabilities for a single $\alpha$. Less aligned fragment distributions resulting from $v$=9,10,11 present tails that are much
more marked than the ones coming from $v$=7 and 8.

The following step for building the experimental observable is the convolution with the detector resolution window
which is taken as a square gate of 0.07 a.u. in kinetic momentum units, corresponding approximately
to a pixel size of 70 $\mu m$.
This as expected, results into the smoothing of very sharp structures such as the peaks associated with $v$=9 and 11 (pannel d).

Finally the theoretical spectrum is obtained as a sum of partial vibrational distributions with weights corresponding to
the vibrational populations as given by Eqs(\ref{eq:pop_7}-\ref{eq:pop_8}).
Panel (e) of figure \ref{fig:intapop} displays the resulting kinetic energy spectrum
which is directly compared with the experimental one. An excellent agreement is obtained, the most noticeable differences
occurring in the vicinity of $v$=9, which corresponds to an energy region particularly sensitive to possible inaccuracies related
with the spatial intensity averaging of the laser.

The information we get from figure \ref{fig:intapop} can be summarized as follows:
apart from the geometrical Abel transformation which is needed to bridge the dissociation probabilities evaluated in
the ($k$-$\theta$) frame, to the photodissociation spectra as recorded on the detector plate,
one has to take into consideration basically two additional facts, when attempting a quantitative interpretation
of the experimental data:

\begin{figure}[!htp]
\begin{center}
   \includegraphics[scale=0.9,angle=0]{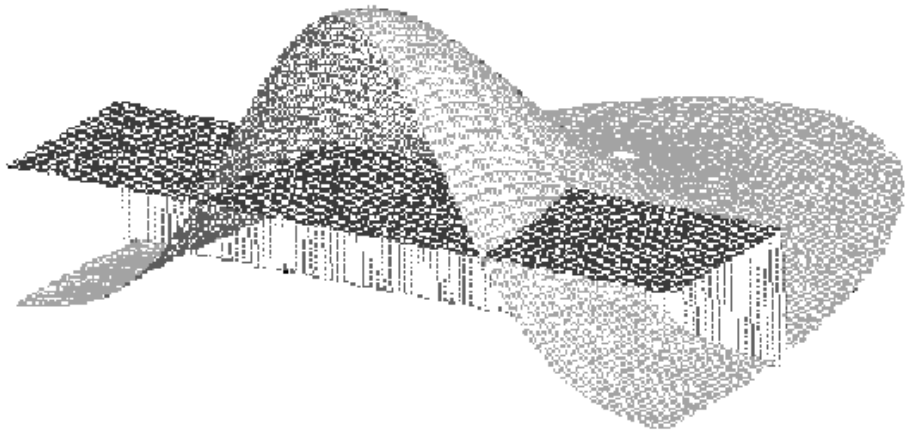}
\end{center}
\caption{\footnotesize The laser field intensity distribution over the molecular beam.}\label{fig:int_beam}
\end{figure}

i) The first is the spatial intensity distribution of the laser. Figure \ref{fig:int_beam}
displays a three-dimensional view of the relative spatial extensions of the laser focal area and that of the
molecular beam as is actually the case in the experiments. Clearly, all the molecules are not subjected to the same intensity
at a given time, requiring thus a spatial averaging, the role of which is one of the most striking.
\begin{figure}[!ht]
\vspace{1cm}
\begin{center}
  \includegraphics[scale=0.5,angle=270]{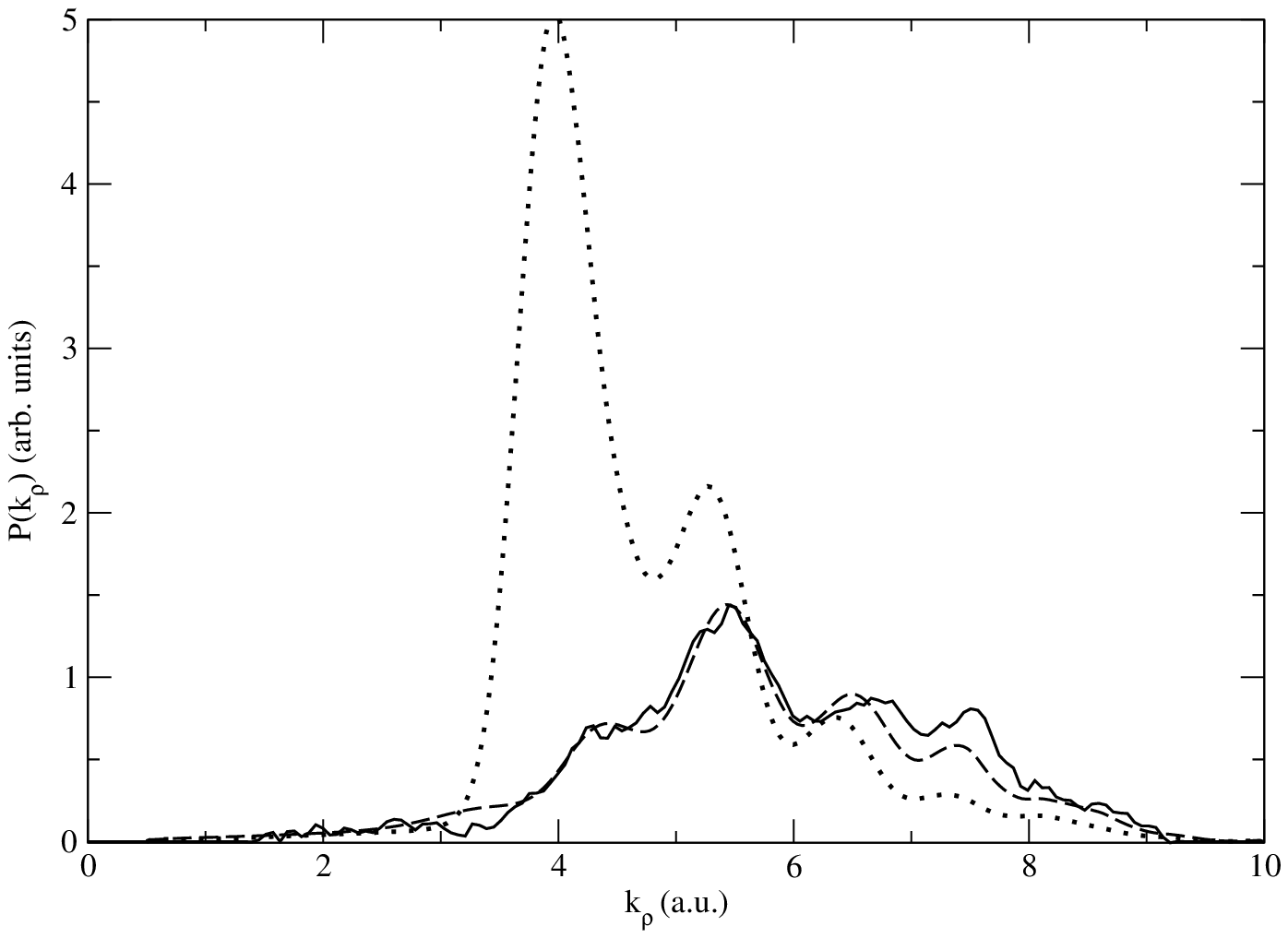}
\end{center}
\caption{\footnotesize The dissociation probabilities calculated with (dashed line) and without (dotted line) averaging on the
laser intensity distribution, versus experimental data (solid line).}\label{fig:int_role}
\end{figure}
Figure \ref{fig:int_role} gathers the spectra on the detector plate, for $\alpha=0$ and as a function of $k_\rho$ for two models:
namely, with and without the spatial averaging over the laser intensity distribution. The results are compared to the
experimentally recorded data. A huge decrease affects the spectrum in the momentum region extending from $k_\rho \simeq$ 3
to 5 a.u. when averaging over the field intensities. This precisely corresponds to the contributions of vibrational levels
$v$=6,7 well protected against potential barriers that are high for lower intensities taking part in the averaging process.
Thus the spatial averaging turns out to be crucial when comparing with experimental spectra.

\begin{figure}[!ht]
\begin{center}
   \includegraphics[scale=0.45,angle=270]{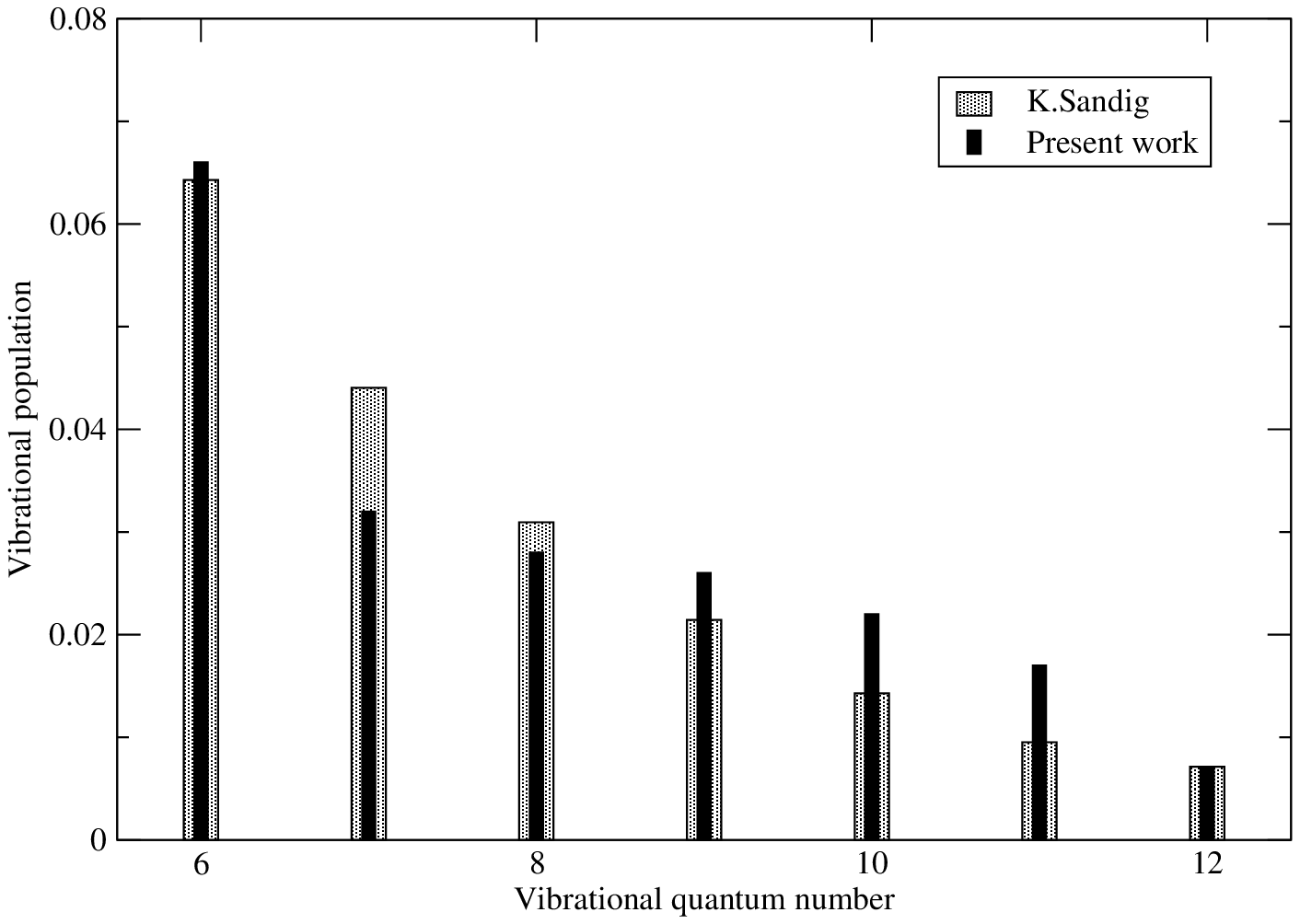}
\end{center}
\caption{\footnotesize Vibrational levels populations as estimated by Sandig in \cite{21} and fitted in the present work.}\label{fig:vib_pops}
\end{figure}

ii) The second is an accurate knowledge of the field-free vibrational populations of the parent ion H$_2^+$, which take part
in Eq.(\ref{eq:pop_7}) through the function $a(v)$. Figure \ref{fig:vib_pops} displays in terms of histograms the relative
vibrational populations of levels $v$=6,...,12 as they result from an estimation based on similar discharge experiments \cite{21}.
They are actually subjected to errors presumably within 10 to 15\% in relative values.
\begin{figure}[!ht]
\begin{center}
  \includegraphics[scale=0.45,angle=270]{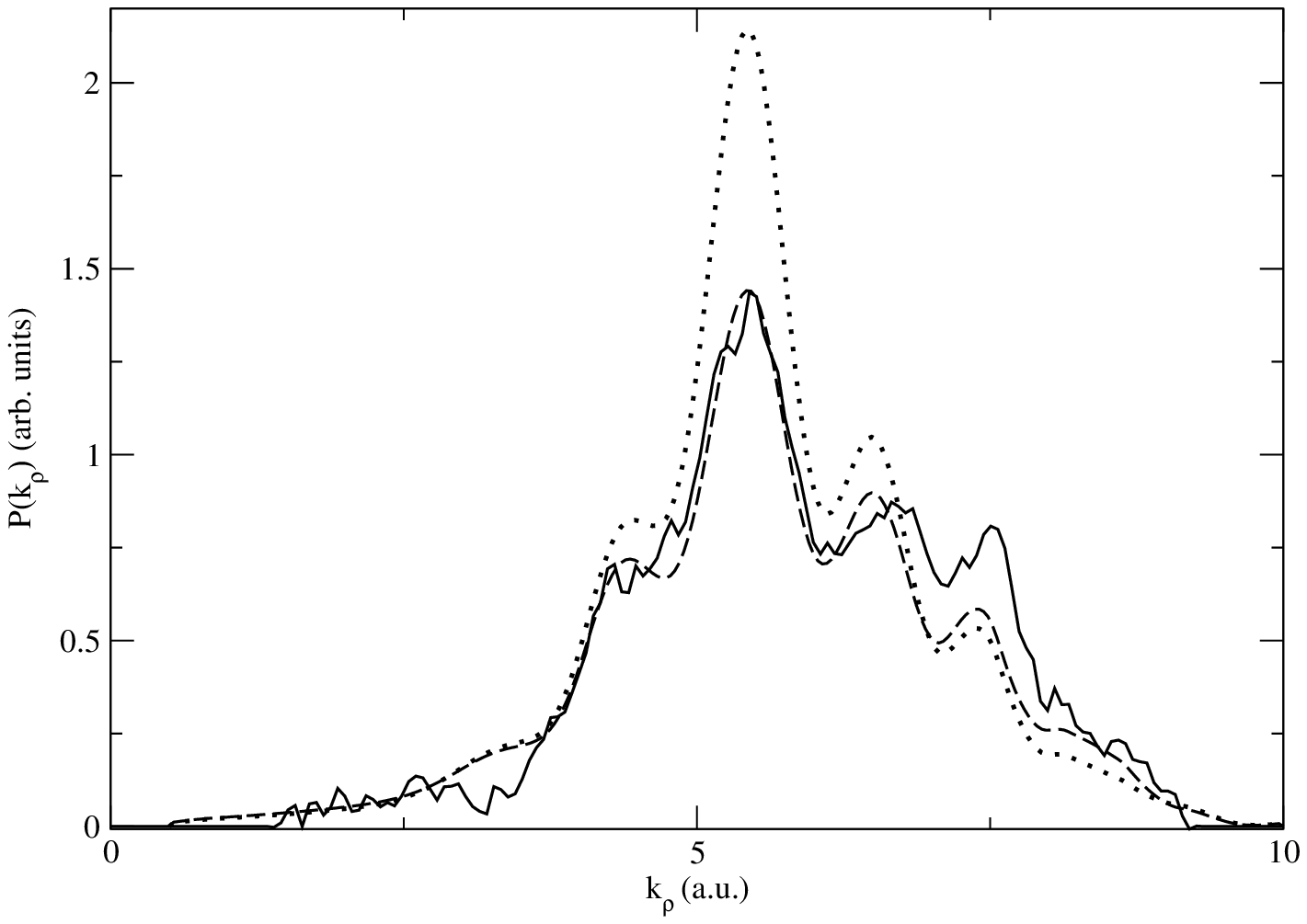}
\end{center}
\caption{\footnotesize Dissociation probabilities calculated using Sandig \cite{21} vibrational levels populations (dotted line)
and the modified ones (dashed line) as compared with the experimental data (solid line).}\label{fig:pops_res}
\end{figure}
A calculation based on them leads to the spectrum illustrated in figure \ref{fig:pops_res} which basically disagrees with the
experimental one over a region close to $k_\rho \simeq $ 5 to 6 a.u., corresponding to the most important peak.
However, a nice agreement is obtained after some small modifications of the vibrational populations, not exceeding
reasonable error bar limits and remaining within the overall decreasing behavior for high $v$'s, as is plotted in figure \ref{fig:vib_pops}.
It is worthwhile noting that this adjustment is done once for all, for given laser parameters (0.7 mJ of total energy)
and is used hereafter for all other theory-versus-experiment comparisons.

\begin{figure}[!htp]
\begin{center}
   \includegraphics[scale=0.7,angle=0]{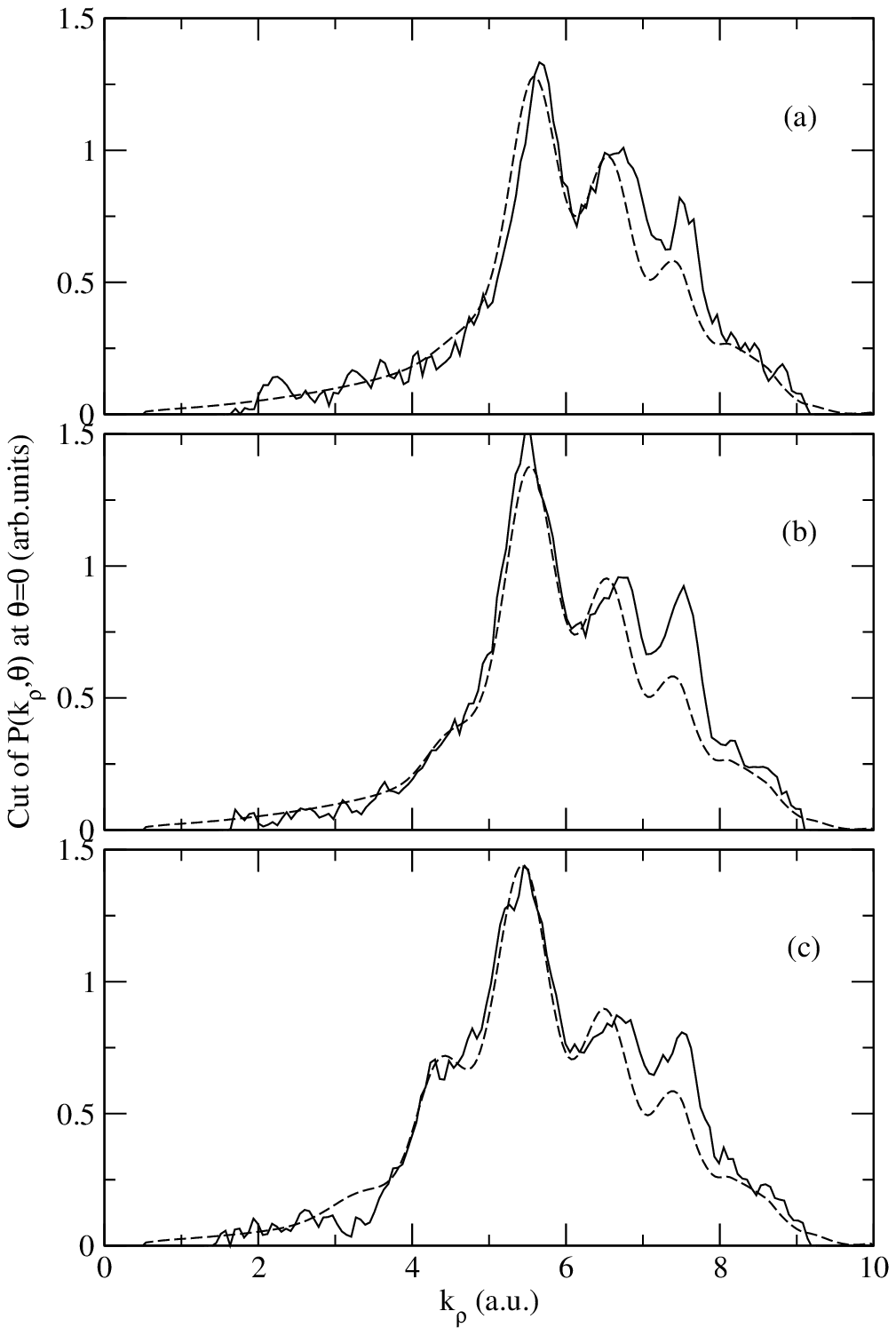}
\end{center}
\caption{\footnotesize Cuts at $\alpha=0$ of the calculated (dashed line) and measured (solid line) kinetic momentum spectra
for: (a) E$_0$=0.3 mJ; (b) E$_0$=0.5 mJ; (c) E$_0$=0.7 mJ.}\label{fig:pk_all}
\end{figure}

The spectra are gathered within the frame of two one-dimensional representations: either as a function of the
kinetic momentum $k_\rho$ or as a function of the angle $\alpha$ on the detector. Figure \ref{fig:pk_all} gives
the cuts (at $\alpha$=0) as a function of $k_\rho$, for three laser fields, whose characteristics are precisely
the ones indicated on Table \ref{tab:exp_params}.
It might be noted here that peak intensity value $I_0$ calculated using Eq.(\ref{eq:int_9}) strongly depends from 
the experimentaly measured parameters $b_x$, $b_y$ and $E_0$. In third column of Table \ref{tab:exp_params}
we give the intensities corresponding to the best agreement
between experimental and calculated spectra 
presented on figure \ref{fig:pk_all}. 
This adjustment is necessary to reproduce correctly the left part of the spectra, which is particularly sensitive to the 
laser intensity, as can be seen in figure \ref{fig:int_role}.
We emphasized that this adjustment has only been performed for one-dimensional spectra corresponding to ($\alpha$=0).
For the same pulse duration the laser intensities are ranging
from low to medium and strong, following the panels a,b, and c. Three features can be emphasized:

i) Three major peaks are obtained, corresponding to the dissociation involving $v$=7, 8 and 9 levels, positioned in this order
in the region $k_\rho \simeq$ 5 to 8 a.u. The theory-experiment agreement is good not only for the positions but also
for the relative heights of these peaks; the most noticeable difference affecting again $v$=9, more sensitive to an accurate
evaluation of the spatial intensity distribution;

ii) The strongest field (E$_0$=0.7 mJ, panel c) reveals the rise of an additional peak at the position of $v$=6.
This is related with the bond softening mechanism, where the radiative coupling induces an important adiabatic barrier
lowering, large enough for the population initially in the bound state $v$=6 to escape through tunneling.
Excellent agreement with experimental results is obtained for this bump in the spectrum ;

iii) The blue tail of the spectrum extending above $k_\rho \simeq$ 8 a.u. corresponds to the photodissociation of initial populations
on $v$=10,11 and 12, which due to possible vibrational trapping effects, is less efficient.
Here again excellent theory-experiment agreement is reached.

\begin{figure}[!htp]
\begin{center}
   \includegraphics[scale=0.7,angle=0]{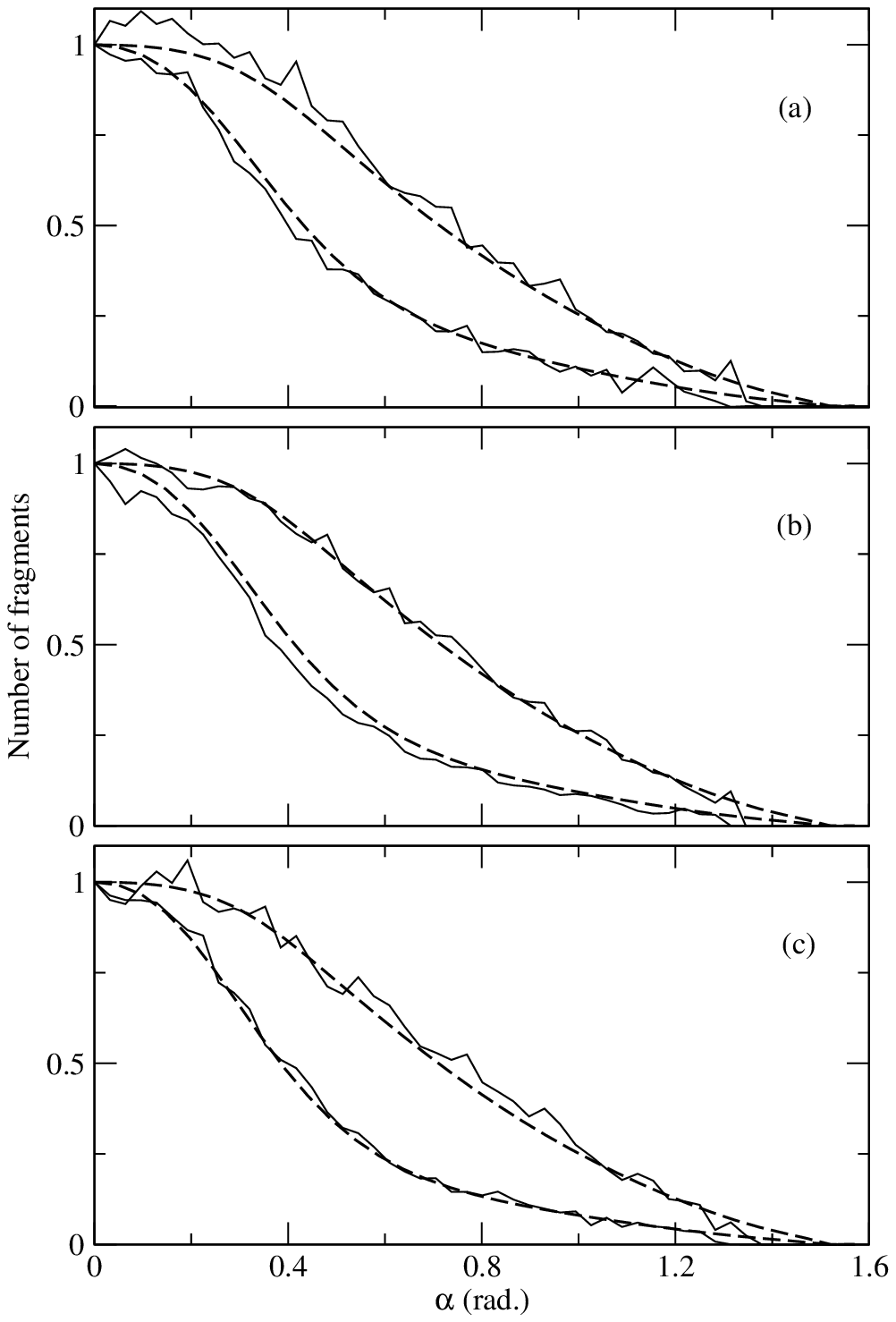}
\end{center}
\caption{\footnotesize Angular distributions of photofragments for: (a) E$_0$=0.3 mJ; (b) E$_0$=0.5 mJ;
(c) E$_0$=0.7 mJ. Solid lines correspond to experimental results, dashed lines to the calculated ones.
(the upper pair to v=7 and the low one to v=8).}\label{fig:pa_all}
\end{figure}

The angular distributions for the same field characteristics, are gathered in figure \ref{fig:pa_all}.
They, precisely, correspond to $\alpha$-dependent 1D representations of fixed $k_\rho$-cuts
of the 3D information of the type displayed in figure \ref{fig:3D_res}. This is done for two different values of
$k_\rho$; namely $k_\rho$=5.5 a.u. and $k_\rho$=6.5 a.u. corresponding to the positions of the maximum amplitudes of the
two major peaks of figure  \ref{fig:pk_all} attributed to $v$=7 and 8, respectively. The following observations can be pointed out:

i) These angular distributions, although labeled as $v$=7 and 8,
actually contain informations originating from $v$=9,10,...,
through the red-tail contributions of these levels (as is clear
from figure \ref{fig:intapop} panel c). The excellent theory-versus-experiment agreement that is reached has to be judged within this
intricate influence of the higher energy part of the spectrum. It
is also worthwhile noting that due to larger experimental errors affecting
$v$=9,10, angular distributions are not studied for higher $v$'s
than 8.

ii) $v$=7 is much better aligned than $v$=8. This is basically due to the bond softening mechanism.
A high potential barrier at $\theta=\pi/2$ protects $v$=7 population against photodissociation. This barrier is lowered at
$\theta$=0 or $\pi$ where the radiative coupling is at its maximum, leading to efficient alignment,
that is even better for increasing intensity. In other words the wavepacket associated with $v$=7 has to skirt
around a high potential barrier at $\theta=\pi/2$ before dissociating, whereas the one associated with
$v$=8 being closer to the top of the barrier can more easily tunnel. The consequence is that dissociation
is facilitated for $\theta$=0, or $\pi$, when the initial population lies on $v$=7.

\begin{figure}[!htp]
\begin{center}
   \includegraphics[scale=0.7,angle=0]{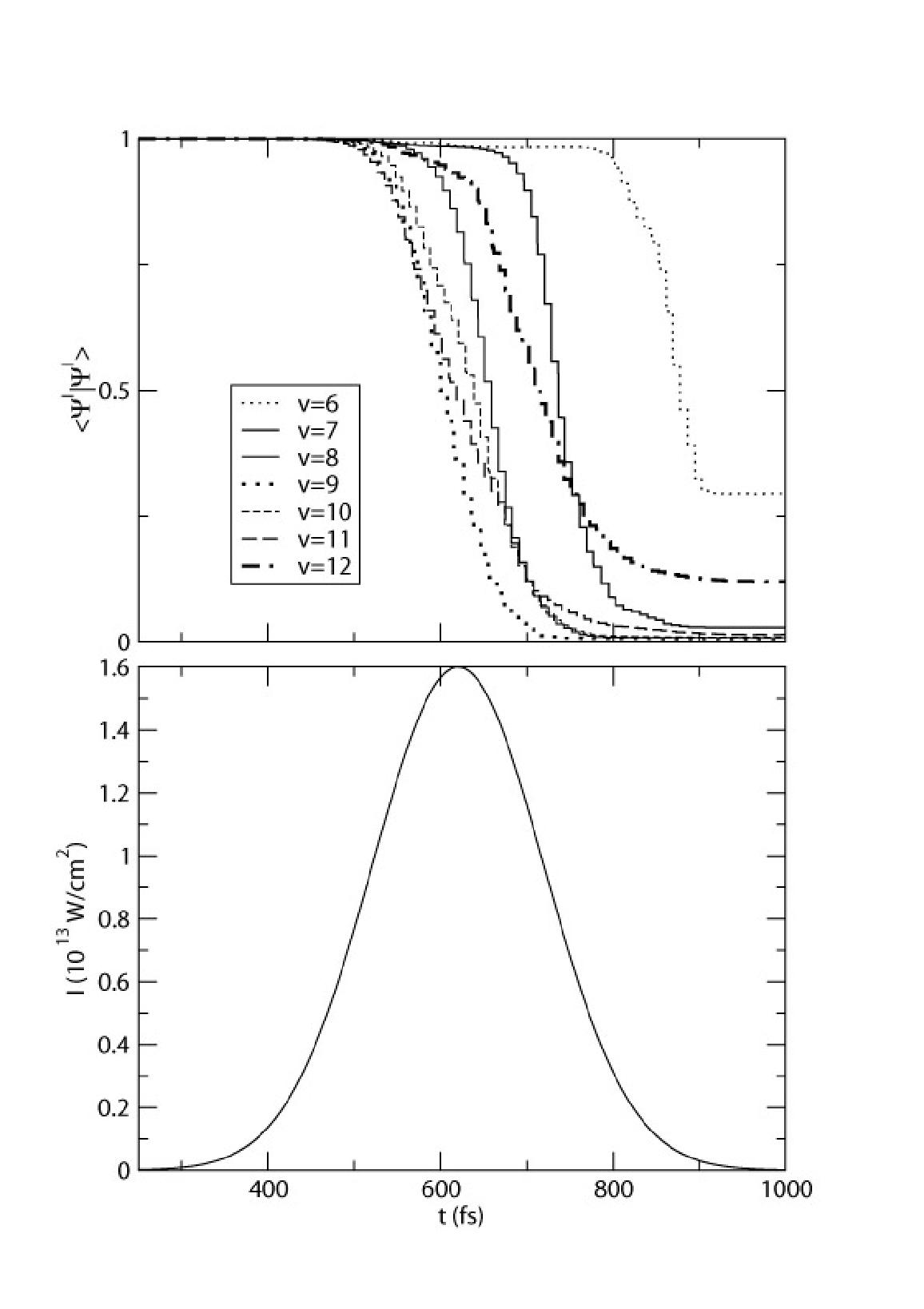}
\end{center}
\caption{\footnotesize Time-resolved decay dynamics of individual vibrational levels in correspondance 
with the temporal laser intensity distribution.}\label{fig:test_norm}
\end{figure}

A better understanding and interpretation of the way following which these complementary bound softening and vibrational
trapping mechanisms ultimately affect the dissociation process, could be gained by a dynamical investigation.
Figure \ref{fig:test_norm} illustrates a time-resolved decay dynamics of individual vibrational levels.
The lower panel gives the temporal shape of the laser intensity for the strongest field into consideration
(E$_0$=0.7 mJ, with the parameters of the last row of Table \ref{tab:exp_params}). The decay dynamics is given
in terms of the decrease of the short range population as a function of time, starting from a given
$v$ of the parent ion H$_2^+$ ({\it i.e.} the time dependence of the norm of the internal region wavepacket $||\Phi^I||^2$, as defined in subsection 2.1).
The behavior of different $v$'s can be classified as follows:

$\bullet$ Levels affected by the bond softening mechanism; namely $v$=6,7,8. The decay starts only after the maximum of the laser pulse,
which is required for the potential barriers to be sufficiently lowered. Although the decay mechanism is rather fast (the slope of
$||\Phi^I||^2$ versus time is large), the photodissociation starting from $v$=6 is not complete, due to the fact that the
potential barrier is closed before total escape towards the asymptotic region.

$\bullet$ $v$=9 which lies at the curve crossing region dissociates completely and faster than all other levels;

$\bullet$ Levels affected by the vibrational trapping mechanism; namely $v$=10,11,12. The populations of these levels start to dissociate
during the laser rise time, but about the maximum intensity their decay rate is lowered (the slope of $||\Phi^I||^2$ versus time
lower than the one of levels $v$=6,7,8). This is basically due to the fact that they are vibrationaly trapped in the temporarily
closed upper adiabatic potential. It is also interesting to note that the population of $v$=12, trapped during the radiative
interaction, partially returns back to the ground state bound potential, in such a way that dissociation starting
from $v$=12 is not complete.

\begin{figure}[!htp]
\begin{center}
   \includegraphics[scale=0.7,angle=0]{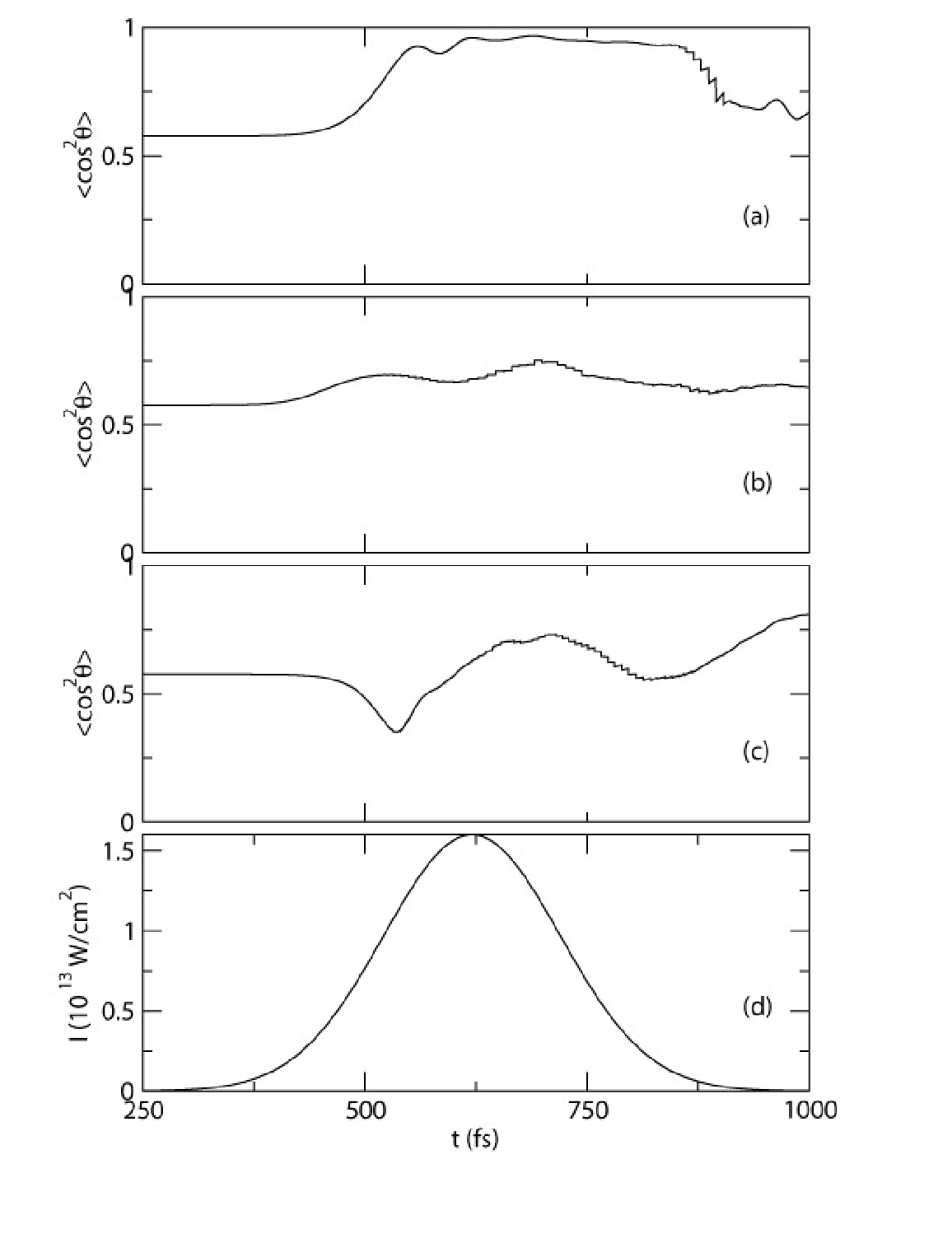}
\end{center}
\caption{\footnotesize Average values of $<\cos^2\theta>$ for the different vibrational levels: (a) v=6; (b) v=9;
(c) v=12 ;(d) - laser temporal profile.}\label{fig:test_cos}
\end{figure}

The dynamical alignment characteristics are illustrated in figure \ref{fig:test_cos}. Here again the lower panel indicates
the temporal shape of the strongest laser. The upper panels display the average value of $\langle\cos^2\theta\rangle$
on the internal region wavepacket;
{\it i.e.} $\langle\Phi^I|\cos^2\theta|\Phi^I\rangle/\langle\Phi^I|\Phi^I\rangle$, indicating the alignment characteristics of the parent ion H$_2^+$ as a
function of time. Three initial levels are in consideration, each pertaining to one of the previously selected classes.
The bond softening mechanism leading to the dissociation of $v$=6 (panel a) results into very efficient alignment during
the pulse, which even remains during the fall off regime. 
\begin{figure}[!ht]
\begin{center}
   \includegraphics[scale=0.9,angle=0]{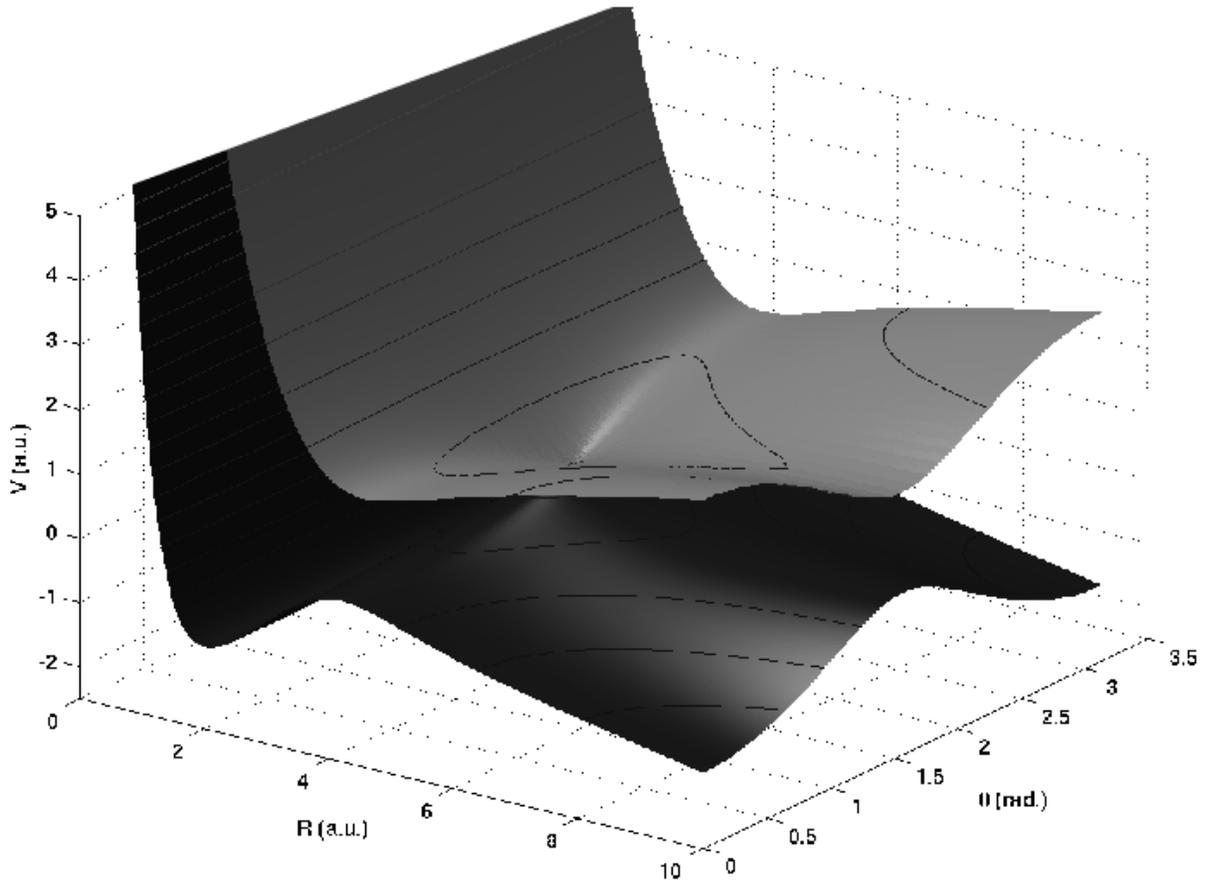}
\end{center}
\caption{\footnotesize 3D adiabatic representation of the H$_2^+$ potential energy surfaces in the presence of
the field with peak intensity $I_0=7.5\times 10^{12}$ W/cm$^2$.}\label{fig:ptls_3D}
\end{figure}
A thorough interpretation, already given in the
literature \cite{24}, can be summarized by referring to a three dimensional representation 
of the adiabatic potential energy surfaces (displayed in figure \ref{fig:ptls_2d}). This is provided
in figure \ref{fig:ptls_3D} for a single photon dressed ground and excited states of H$_2^+$ including
(by diagonalization) the radiative interaction. The photodissociation dynamics starting from $v$=6 is described by a wavepacket that
evolves on the lower adiabatic potential surface. It has first to skirt around the potential barrier at $\theta=\pi/2$
and end up in the lower energy valley at $\theta=0$ or $\pi$. This is why fragments originating from a parent ion in an initial state
well protected by a hardly penetrable potential barrier (as for $v$=6,7,8) are well aligned through the bond softening mechanism.
On the opposite situation are parent ions in an initial state basically pertaining to the upper adiabatic potential energy surface
(as for $v$=10,11,12). This surface presents a minimum around $\theta=\pi/2$, where the population is temporarily trapped.
The dissociation by a single photon absorption further proceeds by a non-adiabatic transition to the lower adiabatic surface.
Such a transition is more efficient for $\theta \simeq \pi/2$ (induced by a lower radiative coupling). Although the last step which is the
evolution on the lower adiabatic surface tends to align the fragments, this effect is less efficient as the parent ion is prepared
at $\theta\simeq\pi/2$ on this surface. The result is clearly understandable in terms of this vibrational trapping  mechanism
for $v$=12 (panel c, figure \ref{fig:test_cos}). During the rise time of the laser pulse, the parent ion is well trapped on the upper
adiabatic surface leading to a misalignment (the bump of the $\langle\cos^2\theta\rangle$ curve at about $t$=550 fs).
There is no noticeable alignment during the whole duration of the pulse. A second misalignment, probably due to the nonadiabatic jump,
is obtained at $t$=800 fs. Finally $v$=9 which is basically not affected neither by bond softening, nor by vibrational trapping,
does not show any alignment characteristics as is clear from panel~b.

\section{Conclusion}

Once the competition between multiphoton ionization and dissociation processes is discarded 
by preparing the parent ion H$_2^+$ through an electric discharge experiment, a rather simple and 
complete quantum modelisation is provided for a theory-versus-experiment comparison of angularly resolved
kinetic energy spectra of photofragments resulting from intense field dissociation.
An Abel transformation relates the probability ${\cal P}(k,\theta)$ for 
a single H$_2^+$ molecule in a given initial ro-vibrational state, to dissociated with kinetic momentum $k$ along its polar direction $\theta$  with respect to the laser polarization vector,
to $P(k_\rho, \alpha)$, the probability for the photofragment H to be detected on a pixel of the detector plate 
labeled by its polar coordinates ($\rho,\alpha$), A quantitative reproduction of experimental data requires some 
statistics over initial ro-vibrational states on one hand and over the spatial distribution of laser intensities
 interacting with molecules positioned at different places in the ionic beam, on the other hand.

An excellent agreement is obtained with experimental spectra and especially for the alignment characteristics 
of the photofragments. A thorough interpretation can be conducted for single vibrational peaks of the spectra 
in terms of basic mechanisms, such as bond softening and vibrational trapping.
The most striking observation is the major role the laser volume effect is playing, in particular 
over lower vibrational levels.

Among our feature prospects, is the elucidation of the role of isotope effects in the photodissociation of 
D$_2^+$ and HD$^+$, which are currently studied in H.Figger's group.

\section*{Acknowledgements}

The authors are indebted to Prof. Hartmut Figger (Max-Planck-Institut f\"ur Quantenoptik, Garching) for 
very fruitful discussions and for communicating them recent and unpublished spectra. 
We acknowledge computation time from Institut du D\'eveloppement 
et des Ressources en Informatique Scientifique (IDRIS, CNRS).

\pagebreak

\renewcommand{\theequation}{A-\arabic{equation}}
  \setcounter{equation}{0}  
  \section*{Appendix}  

\appendix

This appendix is devoted to some geometrical and
and vector relations illustrated in figure \ref{fig:shexp_2}.
$\pmb{r_1}$ and $\pmb{r_2}$ are the vectors pointing H and H$^+$ in the laboratory frame,
$\pmb{R}$ and $\pmb{R_G}$ defined by
\begin{equation}\label{eq:A1}
\pmb{R}=\pmb{r_1}-\pmb{r_2}
\end{equation}
\begin{equation}\label{eq:A2}
\pmb{R_G}=vt\pmb{u_y}=\frac{1}{2} (\pmb{r_1}+\pmb{r_2})
\end{equation}
are the relative internuclear separation and the position of the center of mass $G$, respectively
(by neglecting the contribution of the electron). The right-angle triangles $OO'M$ and $O'M'M$ lead to:
\begin{equation}\label{eq:A3}
\left|\left|\frac{1}{2}\pmb{R}\right|\right|^2=\rho^2+(D-vt)^2
\end{equation}
and
\begin{equation}\label{eq:A4}
||\pmb{r}||^2= \rho^2+D^2
\end{equation}
whereas from the triangle $MGO$ one gets:
\begin{equation}\label{eq:A5}
\frac{1}{2}\pmb{R}=\pmb{r}-\pmb{R_G}.
\end{equation}
The unitary vector $\pmb{u_R}$ along $\pmb{R}$, can be easily evaluated using Eqs(\ref{eq:A2},\ref{eq:A3}) and \mbox{Eq.(\ref{eq:A5})}:
\begin{equation}\label{eq:A6}
\pmb{u_R}=\frac{\pmb{R}}{\left|\left|\pmb{R}\right|\right|}=\frac{\pmb{r}-vt \pmb{u_y}}{\left[\rho^2+(D-vt)^2\right]^{1/2}}
\end{equation}
and its projection over $\pmb{u_y}$ is nothing but:
\begin{equation}\label{eq:A7}
\pmb{u_R}\cdot\pmb{u_y}=\frac{\pmb{r}\pmb{u_y} -vt ||\pmb{u_y}||^2}{\left[\rho^2+(D-vt)^2\right]^{1/2}}= \frac{D-vt}{\left[\rho^2+(D-vt)^2\right]^{1/2}}
\end{equation}
taking into account: $\pmb{r}\cdot\pmb{u_y}=D$.

The polar angles $\theta$ and $\alpha$ positionning H (and $M$) in the center of mass and detector frames
can be related using the right-angle triangle $GM'M$:
\begin{equation}\label{eq:A8}
\cos\theta= \frac{HH'}{||\frac{1}{2}\pmb{R}||}=\frac{\rho\cos\alpha}{||\frac{1}{2}\pmb{R}||}
\end{equation}
or finally taking into account Eq.(\ref{eq:A3}):
\begin{equation}\label{eq:A9}
\cos\theta=\frac{\rho\cos\alpha}{\left[\rho^2+(D-vt)^2\right]^{1/2}}.
\end{equation}

\pagebreak

\end{document}